\documentclass[useAMS,usenatbib,usegraphicx]{mn2e}
\pdfoutput=1
\usepackage{lscape}
\usepackage{color}
\usepackage{longtable}
\usepackage{changepage}
\usepackage{deluxetable}
\title[The morphologies and masses of extremely red galaxies in the Groth Strip survey]{The morphologies and masses of extremely red galaxies in the Groth Strip}
\author[A. Hempel et al.]{A. Hempel,$^{1,3}$\thanks{E-mail: ahempel@iac.es}
D.Crist\'obal-Hornillos,$^{2,5}$
M.Prieto,$^{1,3}$
I.Trujillo,$^{1,3}$ 
M.Balcells,$^{1,3,7}$
\newauthor C. L\'opez-Sanjuan,$^{1,3,6}$ 
D. Abreu,$^{1,3}$
M.C. Eliche-Moral,$^{4}$
L. Dom\'inguez Palmero$^{1,3}$\\
$^{1}$Instituto de Astrof\'isica de Canarias, C./V\'ia L\'actea s/n, La Laguna, 38200, Spain\\
$^{2}$Instituto de Astrof\'isica de Andaluc\'ia (CSIC), Camino Bajo de Huetor 50, Granada, Spain\\
$^{3}$Universidad de La Laguna, C./Francisco S\'anchez s/n, La Laguna, Spain\\
$^{4}$Departamento de Astrof\'isica y Ciencias de la Atm\'osfera, Facultad de C.C. F\'isicas, Universidad Complutense de Madrid, Madrid, Spain\\
$^{5}$Centro de Estudios de F\'isica del Cosmos de Arag\'on, C. General Pizarro, 1-3, 44001 Teruel, Spain\\
$^{6}$Laboratoire d'Astrophysique de Marseille, P$\hat{o}$le de l'Etoile Site de Ch$\hat{a}$teau-Gombert 38, rue Fr\'ed\'eric Joliot-Curie, 13388 Marseille, France\\
$^{7}$ Isaac Newton Group of Telescopes, Apartado de Correos 321, E-38700 Santa Cruz de la Palma, Islas Canarias, Spain}

\begin{document}

\date{Accepted 2011 February 14}

\pagerange{\pageref{firstpage}--\pageref{lastpage}} \pubyear{2011}

\maketitle

\label{firstpage}

\begin{abstract}
We present a new catalogue of EROs from the Groth strip and study the relation between their morphology and mass. With a selection criterion F814W-K$_{s}\ge$4 and  K$_{s}\le$ 21.0 we find 102 EROs, over a survey area of 155 arcmin$^{2}$, leading to a surface density of 0.66 arcmin$^{-2}$. 
The photometric data include $U,B,F606W,F814W,J,K_{s}$ bands. Morphologies are based on a by eye classification and we distinguish between 3 basic classes: compact objects, targets with a disc and/or a bulge component and irregular or merger candidates. An additional group consists of the few objects which could not be classified. \\
The majority of our targets has either a very compact morphology (33$\pm6\%$), or show more or less distinct disc components (41$\pm6\%$). 14$\pm4\%$ are merger or irregulars and 7 objects (approximately 10\%) could not be classified.\\
We also study the dependence of structural parameters (effective radius: r$_\mathrm{eff}$, S\'ersic index: n) on morphological appearance. As expected, EROs that are either compact or show a distinct bulge component have smaller median effective radii (1.22$\pm$0.14 kpc and 3.31$\pm$0.53 kpc) than disc dominated (5.50$\pm$0.51 kpc) or possible irregular galaxies or merger candidates (4.92$\pm$0.14 kpc). More importantly, the S\'ersic index changes from 2.30$\pm$0.34 and 3.24$\pm$0.55, to 1.03$\pm$0.24 and 1.54$\pm$0.40 respectively.\\
As found in previous studies, most the EROs in our sample have redshifts between $z=1$ and $z=2$; however, compact EROs in our sample are found at redshifts as low as $z=0.4$ and as high as $z=2.8$; the latter qualify as well as distant red objects (DRGs). Disc-like EROs are also found up to $z=2.8$; however those with a bulge-disc structure are only seen at $z<1.5$.\\
For each of these EROs we determined the stellar mass and mean population age by fitting synthetic \citet{cb07} spectra to the photometric spectral energy distributions, via $\chi^2$ minimisation. Mass estimates were obtained by assuming an exponentially declining star formation rate with a wide set of parameters, e.g. decay time, redshift of last star formation, metallicity and optical depth.
Total stellar masses for our sample are in the range $9.1<\log(M/M_\odot)<11.6$. We cannot detect significant differences between the stellar mass distribution of the morphological classes. EROs with masses of $\log(M/M_\odot)>11.0$ dominantly show compact morphologies, but also include a significant number of sources with a disc morphology.

\end{abstract}

~\\

\section{Introduction}
It has been 20 years since the first discovery of a population of galaxies with optical to near-infrared (NIR) colours quite different from typical field sources \citep{elston1988,elston1989,hu1994}. Usually defined by {\it R-K}$\gid$5..7 or {\it I-K}$\gid$4..6 colours (in the Vega system), extremely red galaxies are a subset of extremely red objects (EROs) and we will use this more general term throughout this paper.\\
 As diverse as the selection criteria are the stellar populations which produce such red spectral energy distributions. Today, the classification as ERO is beyond the two {\it classic} galaxy types, namely old evolved galaxies with no or very limited recent star 
formation or dusty galaxies with star formation rates (SFR) associated with starbursts, typically with redshift between 1 and 2. The ERO population also comprises normal spiral galaxies \citep{gilbank2003,yan2003,moustakas2004} at slightly lower redshifts. The latter contains a large fraction of edge-on galaxies, where inclination puts a reasonable amount of dust into our line of sight and reddens the SED. Also galaxies harbouring an AGN \citep{alexander2002,brusa2005}, and starburst/AGN combinations \citep{afonso2001} are found among the ERO samples. In addition, several other red galaxy populations have been found, e.g. infrared-detected galaxies \citep{yan2004a}, distant red galaxies (DRGs) \citep{labbe2005,papovich2006} and $BzK$ selected galaxies \citep{daddi2005,hayashi2007}.\\
EROs are relevant because they allow to explore the abundance of massive old ellipticals, which in turn poses a strong test for the two competing scenarios of elliptical galaxy formation: early assembly ($z_\mathrm{f}>$2-3), e.g. by monolithic collapse, and passive luminosity evolution thereafter (PLE models) \citep{tinsley76,pozzetti96}, or hierarchical merging of smaller sized objects \citep{white1978,kauffmann93,somerville01}. Observational evidence has been found for both scenarios: several surveys have detected a deficit of ellipticals at $z >$1, supporting the hierarchical merging models \citep{roche03,kitzbichler06}, while others are consistent with PLE \citep{im02,cimatti02b,somerville04}. \\
In recent years the hierarchical merging scenario in a $\Lambda$CDM universe has been established as the favoured model. Nevertheless, the vast number of different renditions leaves room for dramatically different predictions regarding critical parameters like the number density of massive galaxies at specific times \citep[and references therein]{fontana04,treu05}.\\
 As a whole, extremely red galaxies are among the more massive galaxies (M$\ga10^{11}$M$_{\odot}$) \citep{glazebrook2004,gonzales2008} and the existence of such galaxies with evolved stellar populations at high redshifts is one of the challenges to the hierarchical galaxy formation model \citep{white1978,kauffmann93,somerville01}. However, stellar ages do not necessarily correlate with the build-up of the galaxy mass through merging \citep{delucia2006,trujillo2006, conselice2006,eliche-moral2010}.\\
In the local universe, masses can be determined accurately by studying the dynamical mass of galaxies and obtaining scaling relations such as the fundamental plane for early type galaxies \citep{dressler1987,reda2005,labarbera2008} and dwarf irregulars \citep{vaduvescu2008} and the Tully-Fisher relation for spiral galaxies \citep{tully1977,kassin2007}. Those relations have been used successfully out to z$\sim$1.3, but the most active area of mass assembly, z$>$1.5, is not accessible at current times, although first steps have been taken by e.g. \citet{cenarro2009} and \citet{cappellari2009}. At higher redshifts, the baryonic mass derived from broadband photometry is comparable to the predictions of stellar population models, like \citet{bc03}, \citet{maraston2005} and \citet{cb07}. \\
In this paper we present an analysis of the ERO population in the Groth strip \citep{prieto2005}, based on optical and NIR photometry from the GOYA photometric survey and HST imaging. We describe the morphology, stellar masses and ages, derived from broadband photometry. Specifically we address, first whether all EROs fit in the classic morphological types, and second whether the typical stellar mass of the ERO population changes with cosmic time, and finally, whether at a given redshift the different morphologies of the ERO population correlate with the stellar mass.\\
The paper is structured as follows. In \S \ref{data} we describe briefly the data available from the GOYA survey and the sample selection. \S\ref{morphsec} describes the morphological classification, and \S\ref{massg}  gives an overview how stellar masses were derived using multiband photometry. \S\ref{results} gives details on the morphology and mass estimates for each ERO class.\\
All magnitudes are given in the Vega system and we assume $\Omega_\mathrm{M}$=0.3, $\Omega_{\Lambda}$=0.7 and H$_\mathrm{0}$=70 kms$^{-1}$Mpc$^{-1}$ throughout this work.\\

\section{Data and sample selection}
\label{data}
\begin{figure}
\includegraphics[width=80mm]{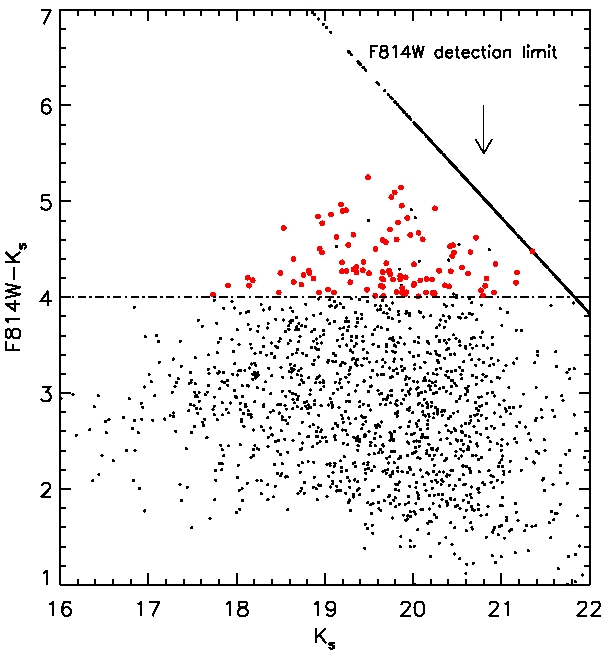}
\caption{Colour-magnitude diagram for the full $K_\mathrm{s}$ selected catalogue (black) in the Groth strip and the final ERO sample (red). The horizontal line represents our colour threshold, the upper envelope is the result of the detection limit in \textit{F814W}. Black dots above the colour threshold show objects which were excluded from our sample during the morphological classification due to their bad image quality.}
\label{KvsIK}
\end{figure}

We have selected a sample of extremely red galaxies (EROs) from the GOYA (Galaxy Origins and Young Assembly) survey \citep{balcells2002}.
This is a combined $K_{s}$ selected catalogue covering the Groth-Westphal strip ($\approx155\mathrm{~arcmin}^{2}$) in 4 optical bands, {\it U,B,F606W} and {\it F814W}, as well as {\it J,K}$_\mathrm{s}$ in the NIR.\\
The $U$ and $B$ imaging were taken with the Wide Field Camera (WFC) at the prime focus of the 2.5m Isaac Newton Telescope (INT); data reduction and catalogue generation are described in \citet{eliche2006} and \citet{domin2008}: 50\% detection efficiencies are 24.8 mag in {\it U} and 25.5 mag in {\it B}.  The \textit{F606W} and \textit{F814W} data originate from the Wide Field and Planetary Camera (WFPC2) on-board \textit{HST}, as part of the original Groth strip survey \citep{groth1994}, that was later analysed for the Medium-Deep Survey (MDS: see, e.g. \citet{ratnatunga1999},  and by the Deep Extragalactic Evolutionary Probe (DEEP; see, e.g., \citet{simard2002}). \\
The NIR $JK_s$ data were obtained with the 1024$\times$1024-pixel INGRID camera at the Cassegrain focus of the 4.2m William Herschel Telescope (WHT).  Data processing and catalogue generation are described in \citet{cristobal2003}. The $50\%$ detection efficiencies range between $\mbox{$K_{\rm s}$ }$ = 21.2 mag and $\mbox{$K_{\rm s}$ }$ = 20.2 mag, depending on the seeing of the individual pointings. \\
The EROs selection is done by running \textsc{SExtractor} on the original ~{\it K}$_\mathrm{s}$-band images, and photometry on all bands, including {\it K}$_\mathrm{s}$, is obtained on apertures of 2.6 arcsec FWHM, in double-image mode, on the images convolved to 1.3 arcsec FWHM. All sources which were classified as stellar objects, based on stellarity greater than 0.8 (given by SExtractor) in the I-band (\textit{F814W}) were excluded from our sample. Figure \ref{KvsIK} shows the colour-magnitude diagram of the whole GOYA catalogue and the final ERO sample. \\

\begin{figure}
\includegraphics[width=80mm]{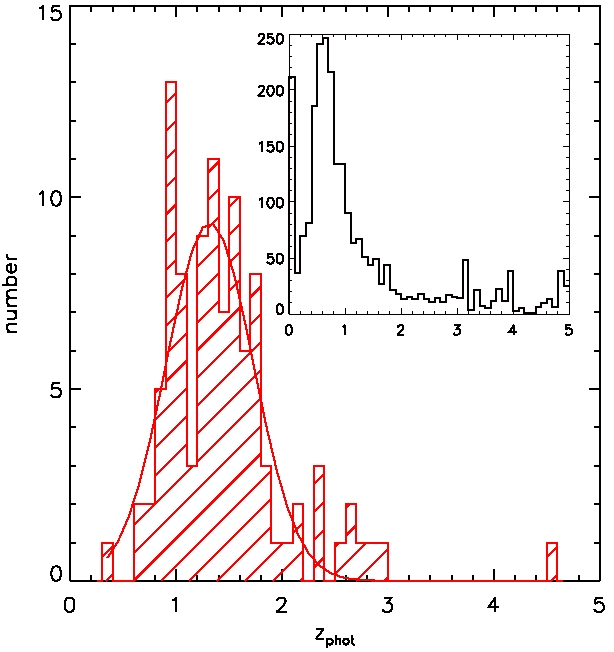}
\caption{Histograms of the distribution of photometric redshifts, $z_\mathrm{phot}$, for our 102 EROs. The inset shows the result for the full $K_\mathrm{s}$ selected catalogue of the Groth strip.}
\label{redshift}
\end{figure}

All our EROs have colours redder than {\it F814W-K}$_\mathrm{s}\ge4.0$. Our initial sample contains 114 objects which we have studied by eye to exclude sources close to the edge of either the \textit{F814W} or {\it K}$_\mathrm{s}$-band image were we have a lower image quality. As result we have excluded 6 objects from our original catalogue. Additional objects have been excluded due to their unresolved morphology (see section \ref{galfit}).\\
Several surveys have obtained spectroscopic redshifts for different galaxy populations in the Groth strip \citep{weiner2005,sarajedini2006}\footnote{http://deep.berkeley.edu}. However, the paucity of emission lines in ERO spectra, plus the lack of prominent emission lines in the visible range at redshifts above 1.4, result in very few EROs having spectroscopically-determined redshifts. We therefor used photometric redshifts available from the GOYA catalogue. These were derived using {\it Hyperz} \citep{bolzonella2000}, redshift errors were estimated from $\sigma_{z_\mathrm{phot}}$= $\sigma_{\delta_\mathrm{z}}\times(1+z_\mathrm{phot})$ with $\sigma_{\delta_\mathrm{z}}$=0.07 \citep{lopez2009}.\\
Based on our photo-z we can see that the used colour threshold of \textit{F814W-K}$_{s}\ge$4 is fairly efficient at selecting galaxies in a redshift range of 1$\le z\le$2. However, this specific colour selection is not very stringent, due to the large variety of star forming histories galaxies may have followed. Figure \ref{redshift} shows the redshift distribution of our final ERO sample, compared to the whole K-selected catalogue. As expected, most EROs have redshifts between 1 and 2, although exceptions at lower and higher redshifts exist. The redshift distribution of our ERO sample peaks at  z=1.32$\pm$0.02, which is in good agreement with \citet{conselice2008} who find an average redshift for an \textit{I-K}$>$4 selected sample of $\langle{z}\rangle$=1.43$\pm$0.32 \citep{conselice2008}.

\section{Morphological classification of EROs}
\label{morphsec}
\begin{figure*}
\includegraphics[width=100mm]{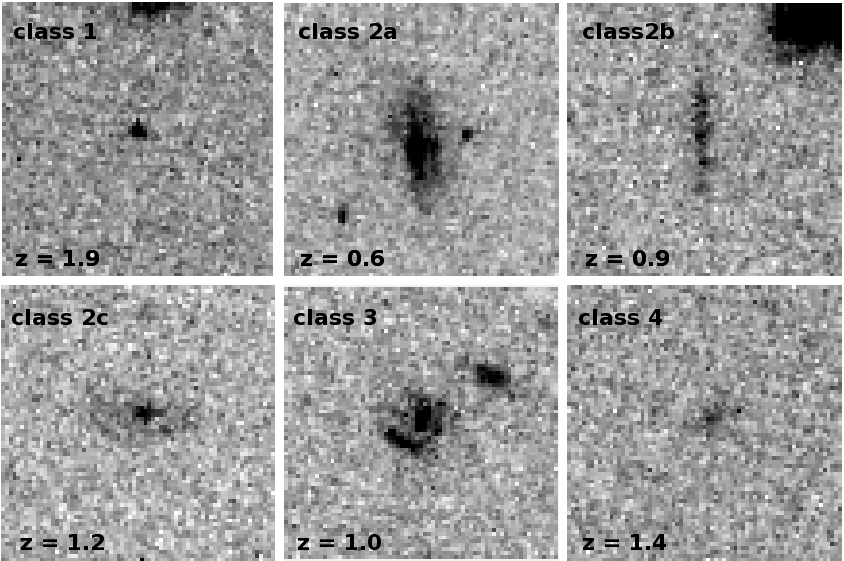}
\caption{HST WFPC2 F814W images of the different morphological classes.  The panels are 7\farcs0 in size, N is up and E left.}
\label{morph}
\end{figure*}

 Morphologies were classified by visual inspection of the \textit{F814W} image of each individual galaxy \citep{yan2003,moustakas2004}. Sometimes down-valued for being subjective, visual classification has a long history and is in fact the method employed to establish the morphological classification of nearby galaxies \citep{devauc1948,ellis2001,desai2007}. We compare this visual classification with the results of a machine-based method based on fitting 2D surface-brightness S\'ersic profiles \citep{roche2002,simard2002,fu2005,stockton2006}, using GALFIT \citep{peng2002}. The later avoid human subjectivity but are also subject to limitations, e.g., when galaxy isophotes are not concentric, aligned ellipses or galaxies with large inclinations. We did not base our classification on automatic determined concentration-asymmetry indices \citep{abraham1996,huertas2008,conselice2008}. Concentration and asymmetry indices are difficult to interpret in EROs which tend to be compact and hence have few pixel where to measure the indices. Furthermore, important systematic errors may occur above z$\sim$1 where the \textit{F814W} filter samples the rest-frame UV.\\
Compared to both methods, a visual inspection can pick up the more subtle morphological details and the sample size is small enough to allow detailed inspection of each of the images.\\
With a mean redshift of 1.32 for the ERO sample, the WFPC2 data sample a rest-frame wavelength of 3500 \AA ~and thus the \textit{F814W} image is sensitive to star formation and to patchy dust extinction. Our morphologies may thus suffer from a 'morphological K-correction bias'. We partially compensate for any such biases by inspecting the ground-based $K$-band image together with the HST/WFPC2 NIR channel.\\
Each galaxy in our sample of 102 EROs was classified individually by four of the authors (A.H., M.P., I.T. and M.B.). Sources for which at least 3 classifications agree are considered as ``secure''. Besides the HST/F814W image, in a few cases we used either the radial profile or the surface brightness isophotes to refine our classification. Initially, we aimed at classifying 3 different morphological types: compact for early type galaxies; extended for disc types; and irregular and merging. However, the second class proved to be quite diverse, containing both galaxies with a bulge and a disc component and galaxies which show no clear bulge component. An examples for each class which show a distinct morphology can be seen in Fig. \ref{morph}.

Finally we differentiate between 6 classes:  
\begin{enumerate}
\item class 1: objects with very compact morphology, like expected for early type galaxies;
\item class 2a: objects with bulge and disc component (early type discs);
\item class 2b: disc galaxies with no clear bulge component, including edge-on discs (late type discs);
\item class 2c: extended objects with bulge and/or disc component, for which no clear classification as either class 2a or class 2b was possible;
\item class 3: irregular or merger candidates;
\item class 4: no clear classification at all.
\end{enumerate}

\begin{figure}
\includegraphics[width=80mm]{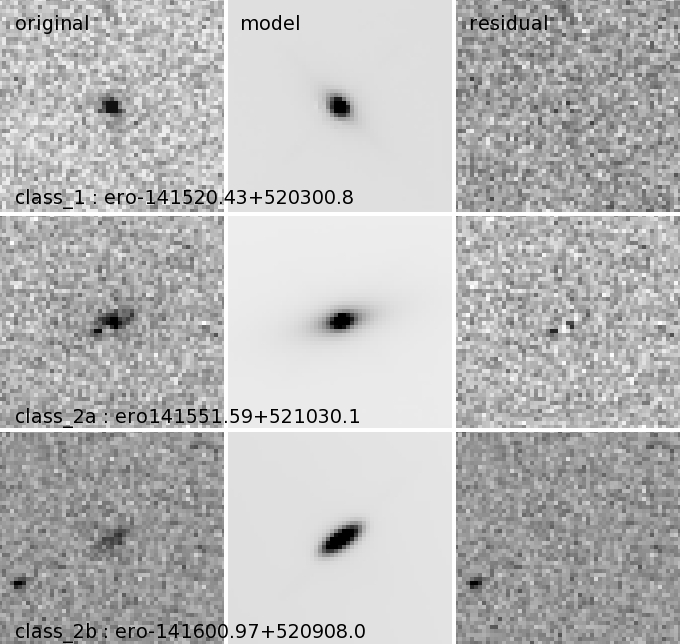}
\caption{Examples of the HST/WFPC2 surface brightness modelling with GALFIT. For one galaxy of class 1, class 2a and class 2b we modelled a single S\'ersic profile (middle column), the panels on the right show the residual image. Each panel is 5.5 arsec in size.}
\label{galfitres}
\end{figure}

We complemented the visual classification with a 2-dimensional surface-brightness fit of each source in our EROs sample, in order to compare the result of both methods, to determine physical sizes and to explore the mass-size correlation between the different morphological classes. \\
The structural parameters were estimated with the GALFIT package \citep{peng2002}, using the HST/WFPC2 F814W. GALFIT simultaneously fits several parameters of an analytic light distribution, thereby minimising $\chi^{2}$, the residual between the original image and the model. As result we can describe the global morphology of our objects in terms of structural parameters, like sizes (given as half-light radius or effective radius $r_\mathrm{eff}$ along semimajor axis a$_\mathrm{e}$) and S\'ersic index n.\\
All our targets were modelled with a S\'ersic profile ($I(r)\propto exp (-(r/r_\mathrm{eff})^{1/n})$). Despite our morphological classification, we keep also the S\'ersic index as free parameter, hence not forcing a pure de Vaucouleurs profile (n=4) on ``elliptical'' EROs or exponential disc profiles (n=1) on objects with bulge and disc components. 
All models were convolved with a PSF obtained from unsaturated stars in the image and extremely bright close-by neighbours were masked. The initial values for the parameters to be fitted were derived by SExtractor. For objects where derived parameters like magnitude, size or position seemed extremely off, the models were tuned by keeping either I$_\mathrm{mag}$ or the position fixed. Figure \ref{galfitres} presents examples of the results, using a single S\'ersic profile.

Results of the morphological classification are presented in paragraph \ref{morphres}. \\

\section{Determining stellar Masses and ages}
\label{massg} 

Stellar masses are estimated by fitting the photometric data to synthetic spectra convolved by the filter transmission function, assuming a known photometric redshift. 
 We use magnitudes corrected for aperture effects and scaled to the best fit spectral energy distribution, which also provides the value for photometric redshift. For 5 objects this procedure produces K-band magnitudes fainter than the detection limit and hence were excluded at this point.\\
 
EROs with a compact morphology are considered to be old evolved systems, whose stellar population formed in a burst like event over  a time span much shorter than their age. As such, population models comprising a single stellar population (SSP) or composite population '$\tau$-models' with rapidly declining SFR ($\propto$ exp(-t/$\tau$)) should be adequate. For morphologically extended EROs, for their similarities with galaxy discs, $\tau$-models with more extended values of $\tau$ should provide a reasonable approximation to their star formation history. The masses presented here are therefore estimated using $\tau$-models. Clearly, the true SFH of these galaxies may be more complex, and we have fitted each galaxy SED with composite populations comprising a SSP and a $\tau$-model. However, considering the small number of available bands, these models suffer from too much degeneracy. \\
The model predictions are based on the models from Charlot and Bruzual \citep{cb07}, and both stellar mass and age for various star formation histories were derived (see section \ref{models}). The term ``age''  indicates the time between the start of the last episode of star formation  ($z_\mathrm{f}$) and the time these galaxies were observed, and therefor represents a ``upper'' limit. 
To simplify, we use ``mass'' as synonym for ``stellar mass''  throughout the paper.

\subsection{Models}
\label{models}
Independent of morphology, we have modelled all EROs with the same set of models and parameters, described in Tab. \ref{param1}. \\
The extinction was modelled with the Calzetti extinction law \citep{calzetti2000}, assuming an average inclination, i.e. orientation effects were not included.

\begin{table}
  \caption{Model parameter for mass estimates. }
 
  \begin{tabular}{ll}
  \hline
   Parameter     &    Range    \\
                 &             \\
 \hline
IMF              &    Salpeter\\
redshift of last SF $^{a}$   &  z$_\mathrm{f}$= 3-8,  $\Delta z$=1\\
age limit  [Gyr]&    0.01 - age$_\mathrm{universe}$ at z\\
exponential decay time [Gyr] &  0.05, 0.1, 0.5, 1.0, 4.0, 7.0\\
metallicity /Z$_{\sun}$     &    0.2, 0.4, 1, 2.5\\
A$_{V}$ (Calzetti extinction law)  &  0.6, 1.0, 2.0, 3.0 \\              
\hline
\label{param1}
\end{tabular}

\medskip
$^{a}$ redshift at which the last episode of extended star formation started.
 \end{table}

The fitting procedure determines which combination of mass and age produces the best fit to the photometric data. 

\begin{figure}
\includegraphics[width=85mm]{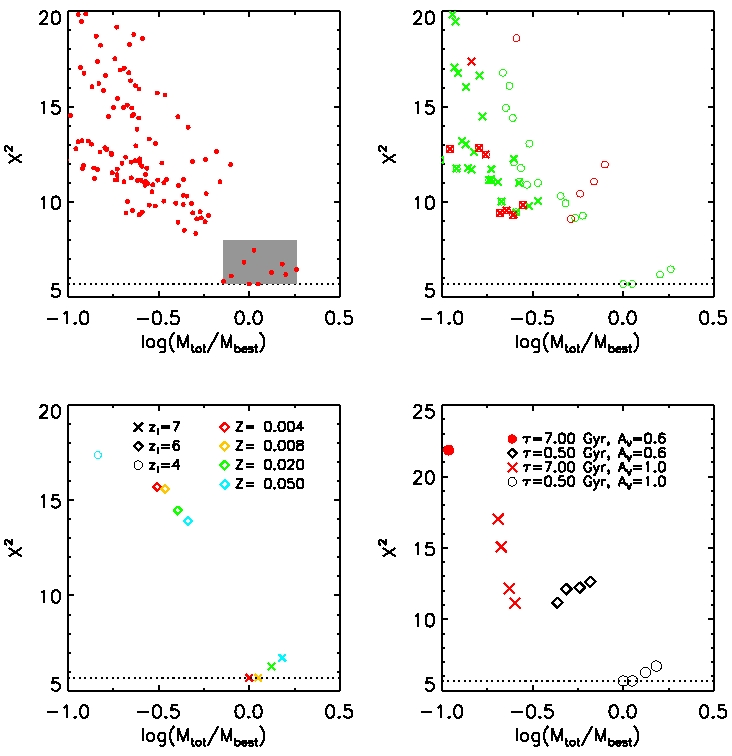}
\caption{Example for the variation of $\chi^{2}$  as result of template fitting (ERO: ero\_141715.09+522142.6). The dotted line represents the minimal $\chi^{2}$ value. The masses have been normalised to the mass of the best solution M$_\mathrm{best}$. {\bf Upper left:} lower $\chi ^{2}$ values for all possible parameter settings. The shaded region indicates the mass range where $\chi_\mathrm{min}^{2}- \chi^{2} \le 2.3$, representing a confidence interval of 68.3\% with 2 degrees of freedom (mass and age of the stellar population). {\bf Upper right:} The symbols represent different formation redshifts: z$_\mathrm{f}$= 4 and 8 (cross, open circles). The colours show different values of A$_\mathrm{v}$:  red: A$_\mathrm{v}$=0.6, green: A$_\mathrm{v}$ = 1.0 and blue: A$_\mathrm{v}$=2.0. The metallicity and exponential decay time cover the whole range. {\bf Lower left}: A$_\mathrm{V}$=0.6, $\tau$ = 0.5 Gyr, z$_\mathrm{f}$= 4,6 and 7 (cross, diamond, open circles). {\bf Lower right}: models with $z_\mathrm{f}$=7, maximal or minimal decay times ($\tau$ = 0.5 Gyr: black), $\tau$ = 7 Gyr: red), metallicity: not restricted. }
\label{chi2}
\end{figure}

From all models (624 models for each ERO) we selected the one with the lowest $\chi ^{2}$ as best fit. Figure \ref{chi2} shows an example for the $\chi^{2}$ values obtained for one specific ERO (ero\_141715.09+522142.6), one of the EROs with a compact morphology.\\

The shaded area (upper left plot) identifies the models with $\chi^{2} -\chi^{2}_{best} \equiv \Delta \chi^2 \le 2.3$, representing a confidence interval of 68.3\% assuming 2 free parameter, stellar mass and age. The majority of models fit the available data poorly, but 10 models result in $\chi^{2}$ values similar to the best solution  (for two pairs of models the $\chi^{2}$ values can not be separated in this plot). The stellar mass for these models ranges from 0.7 $\times M_\mathrm{best}$ up to 1.8 $\times M_\mathrm{best}$. \\
In the remaining three panels we kept some of the model parameters fixed, in order to better separate the influence of specific parameters on $\chi^{2}$. The upper right plot shows the influence of formation redshift (symbols) and dust content (colour), while the other parameters (metallicity and exponential decay time) cover the whole range. For this specific object, models with higher formation redshifts and low extinction show significant better results. Nevertheless, the quality of these models does not improve very much by varying additional parameters like metallicity and exponential decay time. The combination of these parameters might change the stellar mass considerably (as seen in the upper left panel), without improving the quality of already ``good'' fits. \\
In the lower left panel we compare only models with an extinction of A$_\mathrm{v}$=0.6 and $\tau$ = 0.5 Gyr. We can see, that for z$_\mathrm{f}$=6, the metallicity improves the fit quality significantly, while the stellar mass increases only slightly. A formation redshift z$_\mathrm{f}$=7 provides better results in general, however, changing the metallicity does not lead to a lower $\chi^{2}$. In general, the initial metallicity becomes less influential if the object has been formed at earlier times. The lower right panel in this figure shows models with z$_\mathrm{f}$=7, $\tau$ = 0.5, 7.0 Gyr and A$_\mathrm{v}$=0.6, 1.0. All models shown by the same symbol and same colour vary only in their initial metallicity. It becomes clear that for longer exponential decay times, the initial metallicity  becomes more important. The $\chi^{2}$ values for $\tau$ = 0.5 Gyr and A$_\mathrm{v}$=0.6 and 1.0 vary noticeable less than for $\tau$ = 7.0 Gyr.\\
This plot clearly illustrates that all the parameters influence each other in non linear way, e.g. the effect of one parameter on $\chi^{2}$ does not have to be constant, if one of the remaining parameters changes.\\
At the moment we have not calculated the error in age and mass, however, the plot shows that the later can vary by a factor of 2.5 with almost constant $\chi^{2}$, depending on the specific SED template. At the same time, the age for this object will vary by less than 30\%. Such uncertainties in stellar mass agree well with the results of \citet{elsner2008}, who estimated that errors in redshift, $M/L_\mathrm{K}$-ratio, photometry and errors attributed to template fitting can add up to a mean uncertainty of about $\sigma_{\log M} = 0.33~$dex. For an extensive study of stellar mass estimators, their limitations and uncertainties see \citet{longhetti2009}.\\

\section{Results}
\begin{figure*}
\includegraphics[width=18cm]{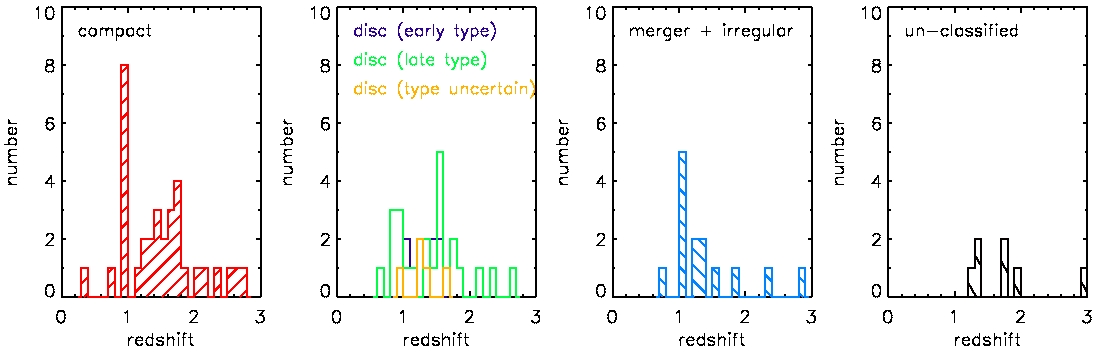}
\caption{Redshift distribution for all classes. The second panel shows the result for all objects with bulge and disc component, disc dominated objects and objects which could not be clearly classified as belonging to either of those.}
\label{classred}
\end{figure*}

\begin{figure}
\includegraphics[width=8.5cm]{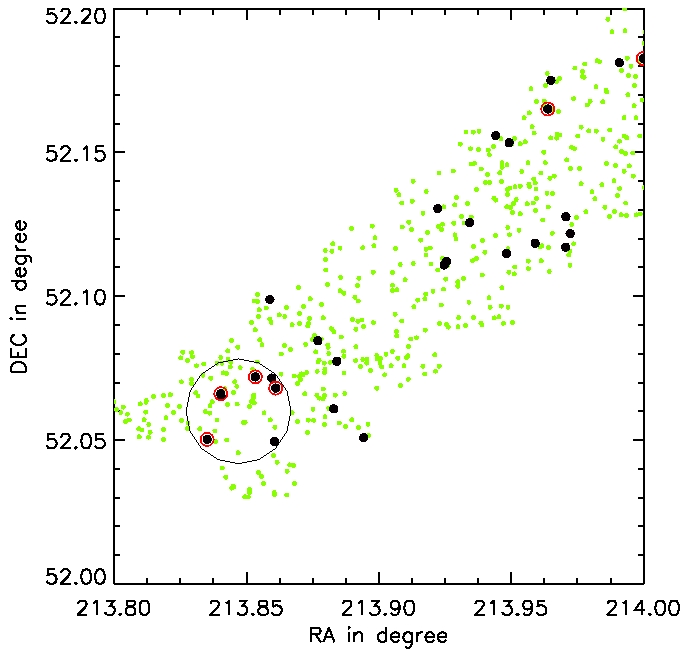}
\caption{Spatial distribution of EROs in a section of Groth strip. The black symbols show all EROs in this section of the field, red circles mark compact objects (class 1) with redshifts of 0.9 $\le z_\mathrm{phot}\le$ 1.0. The black circle with a radius of 1.1 arcmin, corresponds to 0.5 Mpc at redshift 0.95.}
\label{spatial}
\end{figure}

In this section we present our results regarding morphologies, stellar masses, ages and their evolution. Our catalogue is presented in tabular form in the Appendix in Tables A1 and A2, where we list both photometric, morphological, structural and stellar population parameters.
Figure \ref{classred} shows the redshift distribution for each morphological class (see section \ref{morphsec}). The sample of compact objects (class 1, upper left panel in Fig. \ref{massclass} and Fig.\ref{ageclass}) and EROs with bulge and/or disc component (class 2a,b,c, upper right panels in Fig. \ref{massclass} and Fig. \ref{ageclass}) show the clearest redshift distributions, with a peak between redshift 1 and 2. The large number of compact objects with redshifts of 0.9 $\le z_\mathrm{phot}\le$ 1.0 can not be explained by pure low number statistics, hence we looked at the spatial distribution of compact EROs in this region (Figure \ref{spatial}). 4 of the compact objects lie within a field of 1.1 arcmin radius (0.5 Mpc at z=0.95). The large number of such EROs in a very narrow redshift bin hints at an overdensity of compact objects, as we find only a total of 8 such galaxies in the whole $155$ arcmin$^{2}$ field. The resulting surface density is 20 times higher than the average value. However, the 4 EROs lie in an interval $\Delta$z$_\mathrm{phot}$ = z$_\mathrm{max}$-z$_\mathrm{min}$ = 0.07 which, given our typical $z_\mathrm{phot}$ errors, is consistent with zero. In order to proof the physical association of these EROs, spectroscopic redshifts are required. \\

\label{results}

\begin{figure*}
\centering
\includegraphics[width=18cm]{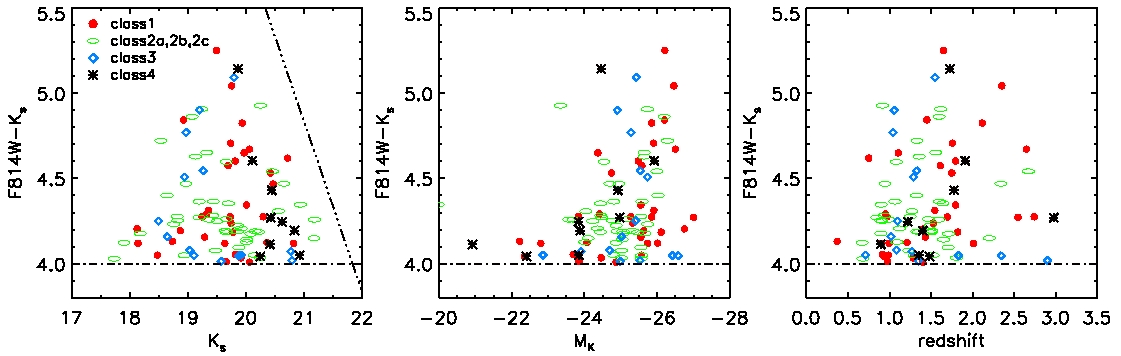}
\caption{Photometric properties of the complete EROs sample: dependency of $F814W-K_\mathrm{s}$ colour on $K_\mathrm{s}$, M$_\mathrm{K}$ and redshift. The same symbols and colours are used in all three panels, the sloped line indicates the colour limit assuming a detection limit of $F814W_\mathrm{lim}$=25.8 }
\label{classIK}
\end{figure*}

In Figure \ref{classIK} we show a summary of the photometric properties (colour $vs$ apparent magnitude ({\it K}$_{s}$), absolute magnitude (M$_{K}$) and redshift) of our EROs sample. We detect EROs from the detection limit of our survey, {\it K}$_\mathrm{s}\sim 21$, up to {\it K}$_\mathrm{s}\approx 18$.  In Fig. \ref{classIK}, left panel, the distribution in colour-apparent magnitude has an apparent triangular shape, such that the reddest EROs are only found at intermediate apparent magnitudes. On the faint side of this distribution, this is due to an observational bias given by the detection limit in $F814W$.  On the bright side, the trend is real. It originates from the fact that the reddest EROs are among the most intrinsically luminous of the sample, as portrayed in the central panel of Figure \ref{classIK}, which shows observed colour against $K$-band absolute magnitude. We detect a mild colour-absolute magnitude relation, but no significant trend of colour with redshift (right panel of Fig. \ref{classIK}).  

\begin{figure*}
\centering
\includegraphics[width=12cm]{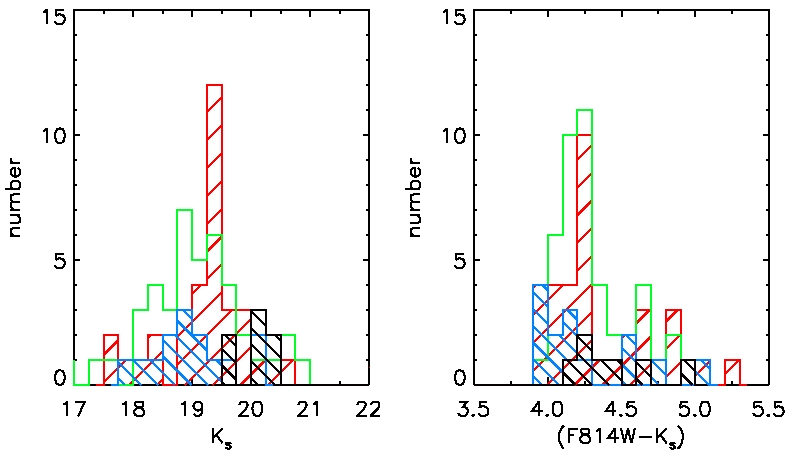}
\caption{K$_\mathrm{s}$-magnitude distribution (left panel) and F814W-K$_\mathrm{s}$ colour distribution (right panel) for our EROs sample, The colours are based on the magnitudes derived from the best fit SED, and not aperture magnitudes. The red histograms represent EROs of class 1, green indicates objects of class 2a,b, and c, i.e. EROs with discs. Irregular galaxies or merging candidates and not classified objects are shown in blue and black.}
\label{morphfrac}
\end{figure*}

All morphological classes are found at all magnitudes and colours, nevertheless, most of the faint objects with irregular of merger characteristics have bluer colours than their brighter counterparts. Brighter EROs (K$_\mathrm{s}\le$19.5) are slightly dominated by disc-like morphologies and EROs with an undetermined morphology are found at the fainter magnitudes (see also Fig. \ref{morphfrac}). 
Based on our data set, we see no clear distinction of the photometric properties between the morphological classes. 

\subsection{Morphology}
\label{morphres}

From the visual classification of our sample we find that 33\%$\pm$6\% of our EROs have a compact morphology and 41\%$\pm$6\%  have a disc component. Irregular galaxies and merger candidates  contribute 14\%$\pm$4\%, while 10\%$\pm$3\% are not classifiable due to low image quality. The uncertainties are quoted solely on the basis of the statistical error of the number of EROs in this morphological class.\\
Our result agrees well with \citet{gilbank2003}, who also find an almost equal fraction of spheroidal/compact EROs and disc-like objects among a sample of 224 EROs (K$<$20,I$_\mathrm{814}$-K)$\ge$ 4.0), 30\%  and 35\%  respectively. 15\% of their EROs show a disturbed/irregular morphology. 
Nevertheless, this result is in contradiction to \citet{yan2003} and \citet{moustakas2004}. The former performed a visual classification of 115 EROs (\textit{F814W-K}$_\mathrm{s}\ge$ 4, 5$\sigma$ median limiting $K_\mathrm{s}$ magnitude of $\approx$ 18.7) into 4 broad categories: spheroids or pure bulge galaxies, bulge-dominated galaxies, disc-like systems with some evidence of a bulge and discs which show no obvious bulge component. Yan and Thompson use the same data set (Medium Deep Survey) as Gilbank et al. and find that approximately 66\% of their EROs are discs or disc dominated and only 34\% have morphologies consistent with bulges or are bulge dominated. These numbers refer to a total of 101 EROs (out of 115) for which the visual classification is either bulge, bulge dominated, discs or disc dominated. However, our results agree in regard to a large fraction of edge-on spirals, 57\%$\pm$11\% of EROs with discs show such an orientation, compared to 40\% in the Yan et al. sample. Moustakas et al. find a combination of 36\% early type galaxies, 55\% late type and 5\% irregulars.\\ 

The most secure classifications of our sample (i.e. at least three out of four classifiers agree) have been obtained for the objects in class 1 (33 of 34), class 2a (13 of 13) and class 2b (24 of 24). The classification for the members of class 2c and class 4 is the least reliable, almost no object in either class (5/5 and 8/9 for class 2c and class 4 respectively) shows a easy to distinguish or unique morphology. For class 3, consisting of irregular and merger candidates, the classification is secure in about 50 \% (8 of 14) of the sample.\\

\subsection{Structural Parameters}
\label{galfit}
\begin{figure*}
\begin{minipage}[t]{0.45\textwidth}
\includegraphics[width=83mm]{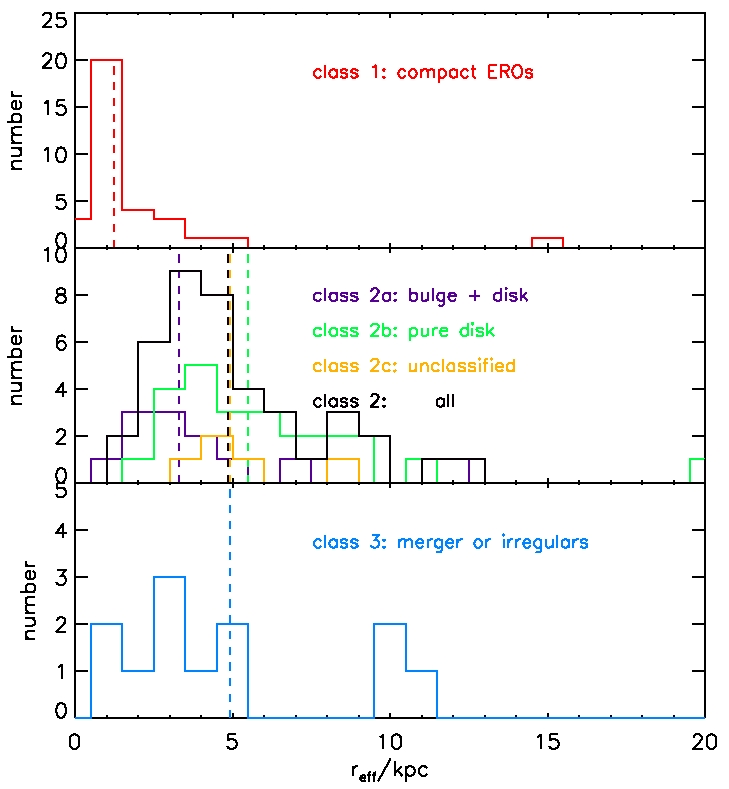}
\caption{Distribution of effective radius (for different morphological classes). The black solid line in the central panel shows the result for the total of class 2. }
\label{radius_histo}
\end{minipage}%
\hspace{1cm}
\begin{minipage}[t]{0.45\textwidth}
\includegraphics[width=83mm]{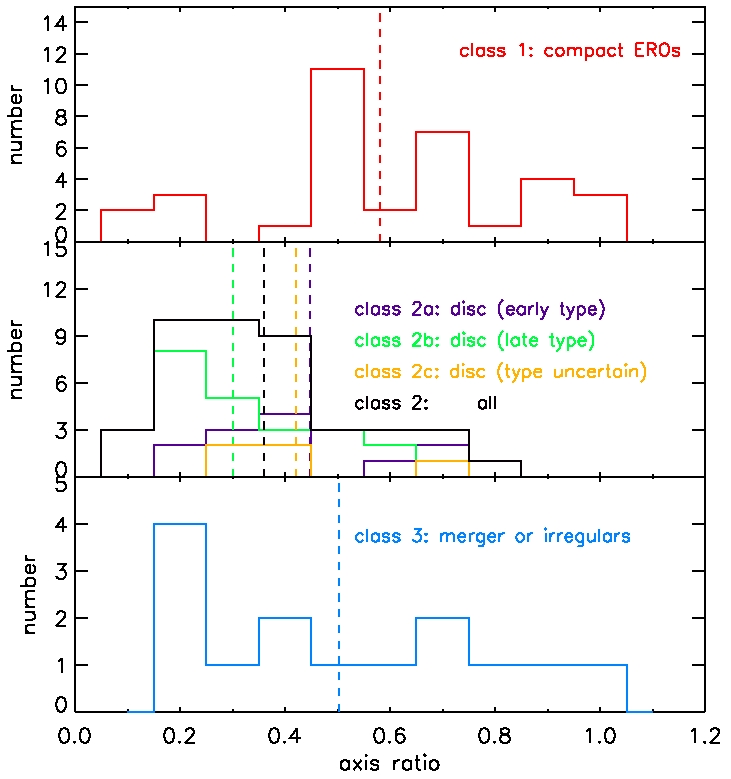}
\caption{Comparison of the axis ratios (derived with GALFIT), for all morphological classes. Colours are assigned to morphological class as in Figure \ref{radius_histo}.}
\label{axisratio}
\end{minipage}
\end{figure*}

In Table A1. we list effective radii, S\'ersic index, axis ratios and model magnitudes, with their errors, for all of the EROs as produced by GALFIT (if not stated otherwise, the effective radii are not circularised), figure \ref{radius_histo} shows the results of the structural analysis for the different morphological classes. For 6 objects with a compact or undetermined morphology (5+1 objects of class 1 and class 4, respectively), the effective radii are extremely small, $\la$ 0.1 kpc. The visual inspection of the original images revealed, that these EROs are barely resolved and although the overall quality of the fit is comparable to the other targets, all parameters show large errors and are very likely faint stars instead of galaxies (these objects have been marked in Tab. \ref{tab1}1).
On this basis we excluded these objects from our analysis.\\
The EROs with compact morphology show the smallest median effective radii (1.19$\pm$0.14 kpc), followed by EROs of mixed morphology (bulge+disc) and disc dominated, 3.31$\pm$0.53 kpc and 5.38$\pm$0.50 kpc respectively. Objects which could belong to either of the last two classes (objects with class 2c morphology) have median sizes of 4.91$\pm$0.14 kpc. Sources which appear irregular or might be part of an ongoing merger and hence show a disturbed morphology have median sizes of about 4.92$\pm$1.06 kpc. The median size of objects with no discernible morphology (class 4) is 2.59$\pm$0.69 kpc.\\
The median S\'ersic index for our compact EROs (class 1), n=2.30$\pm$0.34, is within the errors, in the range for quiescent galaxies, n$\ge$2.5, used by \citet{trujillo2007}, based on the comparison with local galaxies. EROs which appear to be pure discs or at least disc dominated are best fitted with a S\'ersic index of n=1.03$\pm$0.24, compared to a S\'ersic index of 1 for exponential profiles. Objects with bulge and disc components have a median S\'ersic index of n=3.24$\pm$0.55. This implies that the light distribution seems to be dominated by the bulge component, similar to the compact objects in class 1. The result for irregulars or merger candidates (n=1.54$\pm$0.40) lies between compact EROs and disc dominated objects, as we have seen for the effective radius. A median S\'ersic index of n=5.06$\pm$1.61 would imply a very steep surface brightness profile for objects with an unclassified morphology, however, the images have either a low quality and/or the targets are almost unresolved.\\
In our visual classification we consider ``edge-on'' disc dominated as separate morphological class (class 2b) and we would expect that these objects show smaller axis ratios (semi minor axis/semi major axis) than the other morphological types. The plot in figure \ref{axisratio} confirms this expectation, showing an increasing axis ratio from an apparent ``edge-on'' morphology, to ``disc+bulge'' morphologies  and the more compact elliptical EROs (0.3$\pm$0.03, 0.45$\pm$0.06 and 0.62$\pm$0.04 respectively).

Using the masses determined in section \ref{massg}, we show in Figure \ref{mass_radius} the stellar mass-size relation for EROs. The dashed-dotted line shows the relations for early type galaxies (\textit{n}$\geq$2.5) and the dashed line for late type galaxies (\textit{n}$<$2.5), based on the analysis of SDSS galaxies by \citet{shen2003}. This figure shows that compact galaxies deviate more clearly from the local relation than the other morphological types, supporting the strong evolution of the stellar mass-size relation, as described by \citet{trujillo2007} for the most massive galaxies.

\subsection{Stellar Masses}
\label{massres}

\begin{figure*}
\includegraphics[width=130mm]{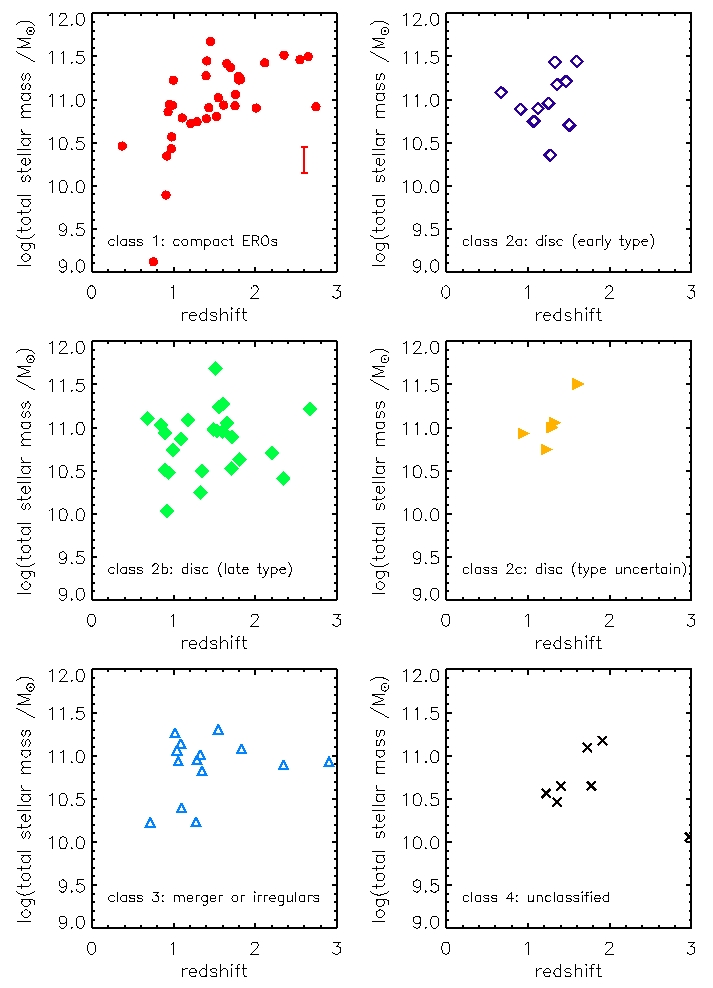}
\caption{Total (stellar) masses for all morphological classes. The error bar in the upper left panel indicates the uncertainties in the mass estimates based on the SED fitting (M$_{max}$/M$_{min}$=2).}
\label{massclass}
\end{figure*}

\begin{figure*}
\includegraphics[width=130mm]{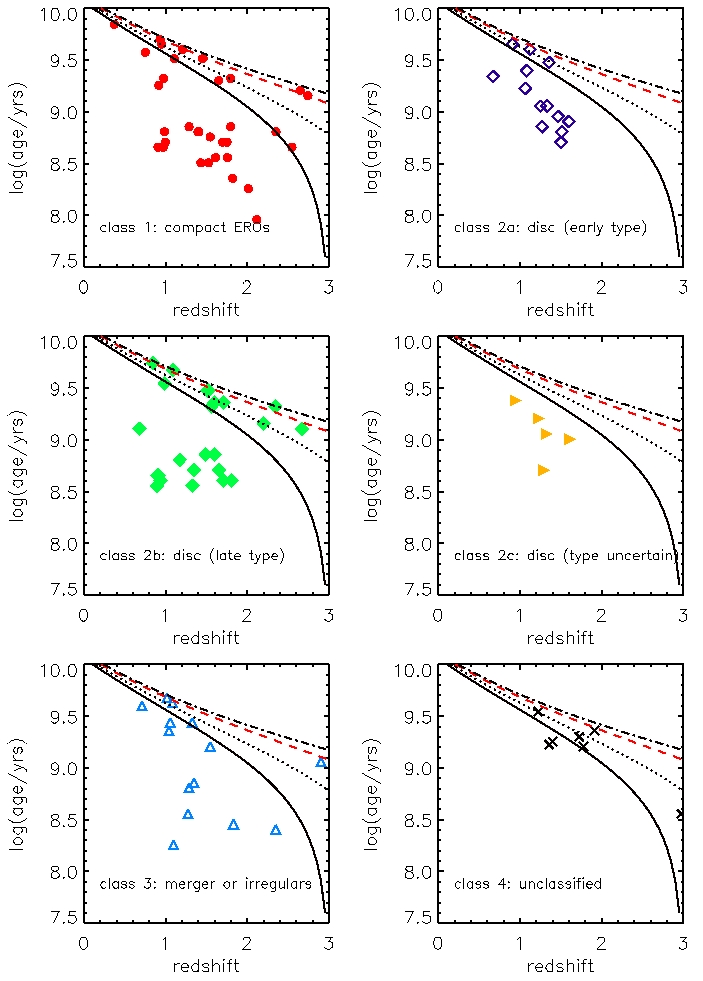}
\caption{Stellar ages based on an exponential declining star formation. The lines correspond to the maximal possible age for a galaxy formed at z$_\mathrm{f}$=8, 6, 4, 3, if observed at the given redshift (from top to bottom, red line corresponds to z$_\mathrm{f}$=6).}
\label{ageclass}
\end{figure*}

\begin{figure}
\includegraphics[width=85mm]{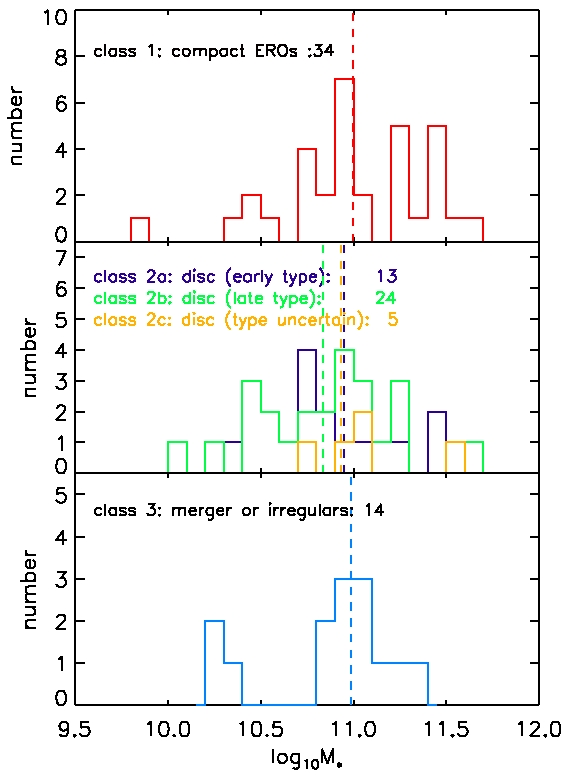}
\caption{Distribution of total stellar mass for distinct morphologies. The vertical lines show the median of each distribution (see Table \ref{mass_med}).}
\label{mass_distrib}
\end{figure}

We now present stellar masses for 97 EROs in our sample, computed as outlined in sec. \ref{massg}. The output of the mass code, namely stellar mass and population age, are shown against redshift in Figures \ref{massclass} and \ref{ageclass}, respectively (see also Table A2.). The vertical error bar in the top-left panel of Figure \ref{massclass} indicates the typical uncertainty of 0.3 dex.  In Figure \ref{ageclass}, where ages are shown against redshift, lines correspond to formation redshifts as detailed in the legend. The median and rms  of the total stellar mass of each morphological class are given in Table \ref{mass_med}, and displayed in Figure \ref{mass_distrib}. Stellar masses for our ERO sample range from $\log(M/M_\odot) = 10.0$ to $\log(M/M_\odot)$=11.8.  Median stellar masses lie between $\log(M/M_\odot) =$ 10.7 and 11.0. The median masses change little from one morphological class to another. Such independence of stellar mass with morphological class is the main result of this paper.  This result was not expected. The common understanding that the most massive galaxies are ellipticals led us to expect EROs with disc or (major) merger morphologies to have significantly lower masses than EROs with compact morphologies. One third of the EROs with a disc dominated morphology (24 EROs in class2b) have masses above $10^{11} M_\odot$ suggests a mechanism to build up massive galaxies that does not involve dissipationless mergers. Such galaxies may have grown through cold accretion \citep{dekel2009a}; alternatively, massive disc-shaped EROs, which must contain vast amounts of dust, are also candidates for remnants of very gas-rich major mergers \citep{hopkins2009}. That EROs include disc-shaped objects with a range of bulge prominence may indicate that the build-up of bulge components in disc galaxies between redshifts 2 and 1 (e.g., \citealt{dekel2009b}) includes important dusty phases.
    

Due to the large scattering in mass between the members of each class and the low numbers, no significant difference in median stellar mass is evident, including the EROs belonging to class 4, whose mean stellar mass ($\log(M/M_\odot) =$10.7) differs by approx. 1$\sigma$ from from the other morphological classes (see Fig. \ref{mass_distrib}). \\
\begin{table*}
 \centering
  \caption{Median of total stellar mass, size (a$_\mathrm{e}$) and S\'ersic index (\it{n}) for each morphological class. The results for r$_\mathrm{eff}$) and S\'ersic index consider only the EROs for which also masses are available. We have excluded objects which diver more than 3 standard deviations from the median value.}
 
  \begin{tabular}{lcccccc}
  \hline
   morphological class      &    log(M$_{\star})$   &   $\sigma$log(M$_{\star})$   & size (a$_\mathrm{e}$)     &  $\sigma$a$_\mathrm{e}$  &  \it{n}         &   $\sigma$n \\
                            &                      &                            & [kpc]                  &      [kpc]                      &           &              \\
 \hline
class 1  : compact           &     10.99           &   0.07                     &     1.22           &     0.14                   &  2.30     &  0.34        \\
class 2a : disc + bulge      &     10.95           &   0.09                     &     3.31           &     0.53                   &  3.24     &  0.55        \\
class 2b : pure (edge-on) disc &   10.93           &   0.08                     &     5.50           &     0.51                   &  1.03     &  0.24        \\
class 2c : disc + bulge (unclassified)& 10.93      &   0.07                     &     4.91           &     0.14                   &  1.72     &  0.40        \\
class 3  : irregular or merger &   10.98           &   0.07                     &     4.92           &     1.06                   &  1.54     &  0.40        \\
class 4  : unclassified      &     10.69           &   0.13                     &     2.59           &     0.69                   &  3.89     &  1.56        \\
\hline
all:                         &     10.66           &    0.04                    &     3.48           &     0.29                   &  2.07     &  0.20        \\
\hline
\label{mass_med}
\end{tabular}
\end{table*}

\begin{figure}
\includegraphics[width=83mm]{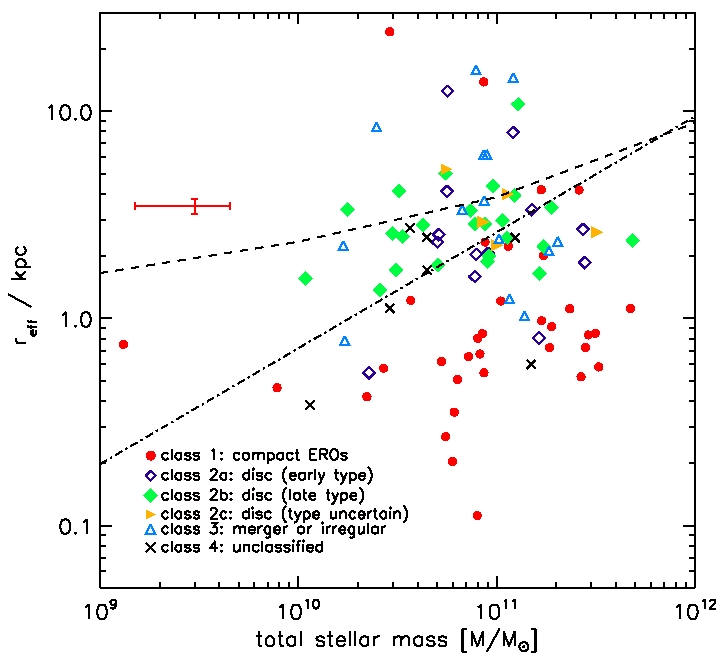}
\caption{The (stellar) mass size distribution for all morphological classes. The effective radii are circularised, $r_\mathrm{eff}$=$a_\mathrm{e}\sqrt{(1-\epsilon)}$, with $\epsilon$ being the ellipticity of the object (1-axis ratio). The dash-dot line shows the local mass size of early type galaxies (n$\geq$2.5) and the dashed line for late type galaxies (n$<$2.5) the Sloan Digital Sky Survey \citep{shen2003}. The stellar mass has been evaluated using a Salpeter IMF. The error bars represent a possible uncertainty in mass by a factor of 2, and the standard deviation for r$_\mathrm{eff}$ (see Table \ref{mass_med}).}
\label{mass_radius}
\end{figure}

Below we discuss each morphological class in more detail.\\
\begin{enumerate}
\item \underline{class 1: compact EROs}\\
The compact morphology of these objects is rather distinct and closely resembles those of an elliptical galaxy with a dominant old stellar population and no or very low ongoing star formation. For most EROs (21 of 34) in this sample the decay times are 100 Myr and less, i.e. the period of star formation is extremely short and on first look resembles a single burst. However, Fig. \ref{agetau} illustrates that for some objects the ratio between t(z$_\mathrm{f}$)-t(z) and $\tau$ is rather small, indicating that these EROs might still form a certain amount of stellar mass. Using a threshold of age/$\tau$=6  \citep{fontana2009} to separate active and quiescent galaxies, 67$\pm$14\% (23/34) of our compact galaxies qualify as quiescent. The fraction of quiescent galaxies among the ``bulge+disc'' galaxies is 84$\pm$25\% (11/13), and drops to 58$\pm$15\% (14/24) for disc dominated galaxies. Combining the galaxies which are either pure bulges (class 1) or show bulge+disc structure (class 2a), we find that approximately 72$\%$ (34/47)  of this sub-sample would qualify as quiescent galaxies. \\

\begin{figure}
\includegraphics[width=8.3cm]{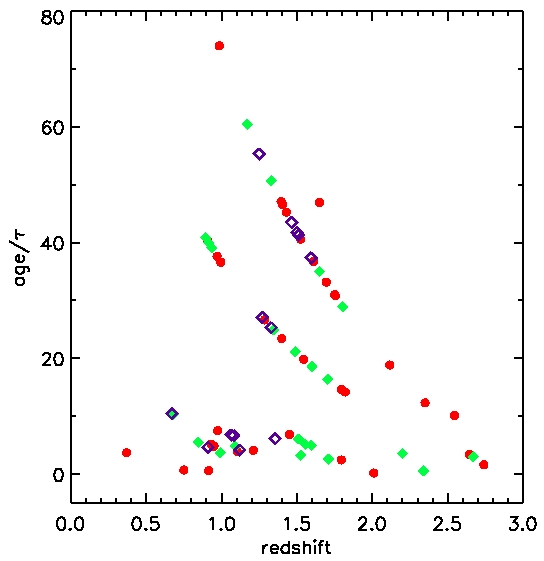}
\caption{Ratio between population age at the time of observation and the exponential decay time $\tau$ against redshift, for EROs with compact morphology (red), bulge+disc (purple) and disc dominated (green).}
\label{agetau}
\end{figure}

Nevertheless, we find that more than 50\% of the entire compact ERO sample have ages of less than 1 Gyr. The youngest object in this class has an apparent age of 0.1 Gyr.\\
The deficiency of low mass objects (M $<$5$\times$10$^{10}$M$_{\odot}$) at higher redshifts is result of the limited depth of our survey. This impression is largely caused by two $z\sim 1$ objects with masses log(M/M$_\odot$)$<$ 10. We have tested the detectability of these two low-mass objects at higher redshifts by simulating their appearance as a function of $z$, assuming passive evolution (Fig.\ref{kvsz}). The simulations show that the lack of low-mass objects at higher redshift is the result of the limited depth of our survey. \\
\begin{figure}
\includegraphics[width=8.3cm]{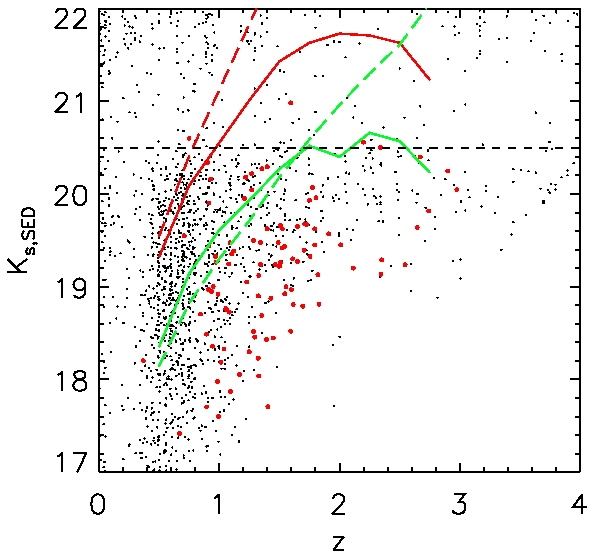}
\caption{$K_{s}$ distribution for galaxies in Groth (black dots) and selected EROs (red dots). The lines are $K_{s}$-band magnitudes derived from evolving the spectral energy distribution of two low mass EROs and the model used to fit these SEDs. The two objects have redshifts of 0.75 (red) and 0.905 (green), and total stellar masses of 0.13$\times10^{10}M_{\odot}$ and 0.78$\times10^{10}M_{\odot}$ respectively. The solid lines represent a SSP model, the dashed line a model with extended star formation. The horizontal line represents a measure of our limiting K-band magnitude. }
\label{kvsz}
\end{figure}

\item \underline{class 2a, 2b, and 2c: EROs with disc component}\\
In this class we find no objects with masses below log(M/M$_{\odot}$)$<$10. Contrary to the compact class, objects with disc-components and with masses between 10.2$<log(M/M_{\sun})<$11.5 are found in the whole redshift range, up to redshift 2.5. This behaviour is clearly visible in panels assigned to class 2 objects in Fig. \ref{massclass}. The mass distributions of the three sub-classes differ due to the limited redshift intervals where objects of each class are found, the mean stellar of all morphological types (with disc-component) agree well within their accuracy (see Tab.\ref{mass_med}).\\
Similar to compact EROs, we find a significant number of objects with stellar populations of more than 1 Gyr of age, both among EROs with and without bulge component. However, no object seems to be younger than 0.3 Gyr, although we would expect these to have lower ages than the compact EROs. For approximately 2/3 of this population, the last star formation event started not later that z$_\mathrm{f}$=3 (see also Fig. \ref{ageclass}).\\ 

\item \underline{class 3: irregular/merging EROs}\\
The majority of objects in this class have masses of approximately 10$^{11}M_{\odot}$, similar to the masses of compact objects or disc-like systems. 
 For none of these objects the last episode of star formation started later than redshift 4, suggesting that population ages of less than 1 Gyr are the result of continuous star formation.  \\

\item \underline{class 4: unclassified}\\
The median stellar mass of this class is slightly lower, 4.6$\times 10^{10}M_{\odot}$.
Like the irregular or merging EROs, the members of this class started forming their stellar mass very early, at redshift 4 and higher. 
With one notable exception, the ERO at redshift 2.98 (ero\_141755.44+522928.5), the stellar population in this group is older than 1 Gyr.

\end{enumerate}
Considering that the mass estimates can vary by factors of 2, we find no dependency of average total mass on the  morphological type of our ERO sample. The stellar mass-size relation (Fig. \ref{mass_radius}) shows a clear dependence on morphological type, with compact galaxies (class 1) having significant smaller sizes than EROs with different morphologies. 

\section*{Summary}
In this paper we have presented a sample of 102 extremely red galaxies with \textit{F814W-K}$_{s}\ge4$ found within the GOYA survey. The EROs morphology was visually classified. We found an almost equal fraction of compact objects (37\%) and objects with a disc component (40\%). Among the later we find a substantial fraction of edge-on spirals, 58\% of all class 2 objects, which is extremely good agreement with the results of \citet{gilbank2003} and \citet{yan2003}. However, there is a discrepancy between the fraction of edge on spirals in respect to the total ERO sample of \citet{yan2003}. \\
Our visual classification agrees well with the results of 2D surface brightness fitting (GALFIT), showing that galaxies which appear compact have both smaller effective radii and larger S\'ersic indices than galaxies with bulge+disc components (larger r$_\mathrm{eff}$ but also larger S\'ersic index). Pure ''disc galaxies'' have both the largest effective radii and the smallest S\'ersic index, while the irregular galaxies or merger candidates seem to represent an intermediate state.\\
Photometric masses were derived by fitting Charlot \& Bruzual \citep{cb07} $\tau$-population models to the SED of the galaxies. Derived stellar masses for our ERO sample range from $\log(M/M_\odot)=10$ to $\log(M/M_\odot)<11.6$. Our median stellar mass is $\log(M/M_\odot) \approx 11.0$. Most notably, the median stellar mass is independent of morphology, arguing against the notion that massive galaxies are spheroids. 
We note that some of the EROs at high redshift (z$>$2), have ages close to the age of the Universe at the redshifts they are observed, which in case of compact objects poses a challenge to the current models of galaxy formation, in which evolved galaxies form at later times through the merger of low mass objects. However, the young objects in class 1 agree well with merger scenarios (see also \citealp{riciardelli2010}).\\
We do not observe a strong evolution within 1$<z<$2 for 10$^{11}M_{\odot}$ mass objects, neither among the compact class nor for EROs with a star forming component. Similar results have been found by \citet{conselice2008} for massive EROs with M$_{\star}> 10^{11}M_{\odot}$ detected in the DEEP2/Palomar fields. The EROs within their K$<$ 19.7 selection have the same upper range of masses at z$\sim$0.8-2.0, therefore indicating little mass growth for this population at this K-band limit. Also we detect a substantial fraction of EROs with total stellar masses below $10^{11}M_{\odot}$ while \citet{conselice2008} state that almost all of their EROs at $K<$19.7 have masses above this value. \\
Looking at the mass-size relations we find that our sample of compact galaxies deviates strongly (circa factors of 4-5) from the local relation, while the galaxies in the remaining morphological classes lie closer to the local mass-size relation for late type galaxies, which is in agreement to previous studies like \citet{trujillo2007}.

\section*{Acknowledgements}
This work was supported by the Consolider-Ingenio 2010 Program grant CSD2006-00070: First Science with the GTC and by grants AYA2006-12955 and AYA2009-11137 from the Spanish Ministry of Science and Innovation.\\
Facilities: HST (WFPC2); ING: William Herschel Telescope (INGRID); ING: Isaac Newton Telescope (WFC)\\

\bibliographystyle{mn2e} 
\bibliography{hempel}

\appendix
\section{Table A1}
Summary of photometric, morphological and structural parameters ((1) target name, (2): photometric redshift, (3): visual classification, (4): security of the vote, (5): aperture magnitude-F814W, (6) aperture magnitude - K$_{s}$, (7): colour, (8): size (a$_\mathrm{e}$ (half-light radius along semimajor axis), (9): S\'ersic index, (10): axis ratio (b/a), (11): F814W magnitude from GALFIT, (12): reduced $\chi^{2}$ by GALFIT). Objects marked with $^{*}$ fulfil also the colour criteria for DRGs: \textit{J-K}$>$ 2.3. The small photometric errors are SExtractor errors only. Objects marked with $^{\#}$ show extremely low effective radii (r$_\mathrm{eff}\leq$0.01 kpc) and were excluded from our final target list.
\newpage
\clearpage
\begin{deluxetable}{lcrrcccccccc}
\tabletypesize{\footnotesize}
\rotate
\tablewidth{0pt}

\tablecaption{Summary of photometric, morphological and structural parameters. For an explanation of the columns please see the text in this section.}
\label{tab1}

\tablehead{

\colhead{Name} & \colhead{z$_\mathrm{phot}$} & \colhead{class} &\colhead{vote} & \colhead{F814W} &\colhead{K$_{s}$} & \colhead{F814W-K$_{s}$} &\colhead{a$_\mathrm{e}$} & \colhead{n} & \colhead{axis ratio} & \colhead{F814W$_\mathrm{GALFIT}$} &  \colhead{$\chi^{2}_\mathrm{GALFIT}$}\\
\colhead{}     & \colhead{}                 & \colhead{}      &\colhead{}     &\colhead{/mag}    &\colhead{/mag}    & \colhead{/mag}           &\colhead{/kpc}             &\colhead{}   & \colhead{}           & \colhead{/mag}                    & \colhead{}     \\
\colhead{(1)}  & \colhead{(2)}              & \colhead{(3)}   &\colhead{(4)}  &\colhead{(5)}    &\colhead{(6)}    & \colhead{(7)}           &\colhead{(8)}             &\colhead{(9)}& \colhead{(10)}       &\colhead{(11)}  &\colhead{(12)}  }
\startdata
ero\_141520.43+520300.8  &  0.92$\pm$0.13  &   1  &     secure  &  23.87$\pm$ 0.08  &  19.82$\pm$ 0.14  &   4.05  &   0.69$\pm$ 0.17  &    6.86$\pm$   3.99  &  0.37$\pm$   0.12  &  23.50$\pm$  0.08  &   0.720 \\
ero\_141521.67+520358.0  &  0.97$\pm$0.14  &   1  &     secure  &  23.67$\pm$ 0.03  &  19.66$\pm$ 0.89  &   4.01  &   1.32$\pm$ 0.18  &    1.75$\pm$   0.63  &  0.85$\pm$   0.16  &  24.13$\pm$  0.00  &   0.720 \\
ero\_141521.72+520354.2  &  1.04$\pm$0.14  &   3  &  un-secure  &  23.74$\pm$ 0.01  &  18.97$\pm$ 0.01  &   4.77  &   1.34$\pm$ 0.19  &    1.75$\pm$   0.63  &  0.85$\pm$   0.16  &  24.13$\pm$  0.00  &   0.720 \\
ero\_141524.80+520419.0  &  0.95$\pm$0.14  &   1  &     secure  &  23.61$\pm$ 0.03  &  19.32$\pm$ 0.03  &   4.29  &   2.76$\pm$ 0.35  &    1.70$\pm$   0.38  &  0.71$\pm$   0.07  &  23.36$\pm$  0.08  &   0.716 \\
ero\_141526.12+520555.9  &  1.61$\pm$0.18  &  2c  &  un-secure  &  24.15$\pm$ 0.02  &  19.24$\pm$ 0.03  &   4.91  &   4.66$\pm$ 1.02  &    1.53$\pm$   0.49  &  0.31$\pm$   0.05  &  23.97$\pm$  0.14  &   0.709 \\
ero\_141526.29+520417.5  &  2.34$\pm$0.23  &  2b  &     secure  &  25.12$\pm$ 0.01  &  20.65$\pm$ 0.53  &   4.47  &   3.56$\pm$ 0.35  &    0.33$\pm$   0.25  &  0.15$\pm$   0.05  &  24.48$\pm$  0.10  &   0.734 \\
ero\_141526.54+520258.1  &  1.75$\pm$0.19  &   1  &     secure  &  24.95$\pm$ 0.02  &  20.42$\pm$ 0.95  &   4.53  &   1.72$\pm$ 0.41  &    2.58$\pm$   1.81  &  0.24$\pm$   0.13  &  24.46$\pm$  0.12  &   0.733 \\
ero\_141526.65+520405.0  &  0.91$\pm$0.13  &   1  &     secure  &  24.94$\pm$ 0.02  &  20.82$\pm$ 0.27  &   4.12  &   0.54$\pm$ 0.14  &    0.13$\pm$   2.51  &  0.75$\pm$   0.36  &  24.95$\pm$  0.08  &   0.755 \\
ero\_141528.89+520415.9$^{\#}$  &  2.00$\pm$0.21  &   1  &     secure  &  24.49$\pm$ 0.46  &  20.46$\pm$ 0.51  &   4.03  &   0.01$\pm$ 0.56  &    9.09$\pm$ 174.96  &  0.99$\pm$   3.93  &  24.16$\pm$ 12.95  &   0.742 \\
ero\_141530.47+520504.3  &  0.68$\pm$0.12  &  2b  &     secure  &  21.76$\pm$ 0.00  &  17.73$\pm$ 0.03  &   4.03  &  20.17$\pm$ 1.88  &    1.50$\pm$   0.13  &  0.29$\pm$   0.01  &  20.55$\pm$  0.08  &   1.591 \\
ero\_141531.91+520339.0  &  1.52$\pm$0.18  &   1  &  un-secure  &  23.98$\pm$ 0.37  &  19.74$\pm$ 0.61  &   4.24  &   0.51$\pm$ 0.07  &    3.29$\pm$   1.66  &  0.98$\pm$   0.22  &  23.55$\pm$  0.04  &   0.715 \\
ero\_141532.19+520438.6  &  1.40$\pm$0.17  &   1  &     secure  &  24.07$\pm$ 0.03  &  20.06$\pm$ 0.13  &   4.01  &   0.44$\pm$ 0.26  &   12.81$\pm$  17.93  &  0.21$\pm$   0.14  &  23.84$\pm$  0.38  &   0.718 \\
ero\_141534.63+520303.0  &  2.67$\pm$0.26  &  2b  &     secure  &  24.99$\pm$ 0.01  &  20.45$\pm$ 0.26  &   4.54  &   2.60$\pm$ 0.61  &    0.07$\pm$   0.60  &  0.40$\pm$   0.14  &  25.07$\pm$  0.25  &   1.959 \\
ero\_141541.34+520749.7  &  1.09$\pm$0.15  &   3  &     secure  &  22.75$\pm$ 0.02  &  18.50$\pm$ 0.02  &   4.25  &  10.38$\pm$ 1.68  &    2.02$\pm$   0.28  &  0.66$\pm$   0.04  &  21.88$\pm$  0.13  &   0.744 \\
ero\_141541.93+520639.0  &  1.70$\pm$0.19  &   1  &     secure  &  23.52$\pm$ 0.03  &  19.24$\pm$ 0.43  &   4.27  &   1.64$\pm$ 0.15  &    1.58$\pm$   0.47  &  0.46$\pm$   0.06  &  23.48$\pm$  0.04  &   0.730 \\
ero\_141542.14+520643.8  &  0.94$\pm$0.14  &  2b  &     secure  &  23.71$\pm$ 0.01  &  19.43$\pm$ 0.01  &   4.28  &   8.26$\pm$ 0.57  &    0.15$\pm$   0.08  &  0.10$\pm$   0.04  &  23.71$\pm$  0.16  &   1.096 \\
ero\_141544.23+520731.9  &  1.09$\pm$0.15  &  2b  &     secure  &  24.13$\pm$ 0.01  &  20.00$\pm$ 0.01  &   4.13  &   8.34$\pm$10.24  &    0.04$\pm$   0.12  &  0.16$\pm$   0.04  &  24.50$\pm$  0.00  &   0.776 \\
ero\_141546.60+520921.1  &  1.77$\pm$0.19  &   4  &  un-secure  &  24.87$\pm$ 0.01  &  20.44$\pm$ 0.78  &   4.43  &   3.19$\pm$ 6.06  &    0.05$\pm$   0.48  &  0.29$\pm$   0.08  &  24.91$\pm$  0.13  &   0.718 \\
ero\_141547.57+520653.5  &  1.52$\pm$0.18  &  2b  &     secure  &  24.20$\pm$ 0.02  &  20.15$\pm$ 0.87  &   4.05  &   3.03$\pm$ 0.74  &    2.37$\pm$   0.99  &  0.44$\pm$   0.09  &  23.88$\pm$  0.15  &   0.740 \\
ero\_141547.81+520912.0$^{*}$   &  1.83$\pm$0.20  &   3  &  un-secure  &  23.15$\pm$ 0.03  &  19.10$\pm$ 0.07  &   4.05  &  30.80$\pm$94.13  &   19.73$\pm$  18.08  &  0.22$\pm$   0.04  &  22.36$\pm$  0.92  &   0.764 \\
ero\_141550.16+520706.3  &  1.60$\pm$0.18  &  2b  &     secure  &  23.80$\pm$ 0.01  &  19.44$\pm$ 0.42  &   4.37  &   4.46$\pm$ 0.42  &    0.74$\pm$   0.16  &  0.59$\pm$   0.04  &  23.44$\pm$  0.06  &   0.732 \\
ero\_141551.31+520954.3  &  0.97$\pm$0.14  &   1  &     secure  &  24.00$\pm$ 0.03  &  19.73$\pm$ 0.04  &   4.28  &   0.71$\pm$ 0.19  &    5.83$\pm$   4.20  &  0.66$\pm$   0.17  &  23.73$\pm$  0.15  &   0.723 \\
ero\_141551.59+521030.1  &  1.06$\pm$0.14  &  2a  &     secure  &  23.56$\pm$ 0.03  &  19.20$\pm$ 0.02  &   4.36  &   7.25$\pm$ 4.61  &    6.28$\pm$   2.62  &  0.32$\pm$   0.05  &  22.88$\pm$  0.29  &   0.727 \\
ero\_141552.93+520701.4  &  1.54$\pm$0.18  &   1  &     secure  &  23.67$\pm$ 0.03  &  19.36$\pm$ 0.54  &   4.31  &   1.75$\pm$ 0.20  &    2.36$\pm$   0.74  &  0.48$\pm$   0.07  &  23.42$\pm$  0.06  &   0.736 \\
ero\_141552.95+520739.5  &  1.79$\pm$0.20  &   1  &     secure  &  24.35$\pm$ 0.03  &  20.01$\pm$ 0.16  &   4.34  &   0.86$\pm$ 0.22  &    3.80$\pm$   2.79  &  0.70$\pm$   0.17  &  24.00$\pm$  0.13  &   0.764 \\
ero\_141553.33+520718.4  &  1.32$\pm$0.16  &   3  &     secure  &  23.81$\pm$ 0.01  &  19.27$\pm$ 0.02  &   4.55  &   3.08$\pm$ 1.84  &    0.05$\pm$   0.21  &  0.62$\pm$   0.08  &  24.09$\pm$  0.09  &   0.808 \\
ero\_141557.80+521052.2  &  0.90$\pm$0.13  &  2b  &     secure  &  23.08$\pm$ 0.00  &  18.82$\pm$ 0.37  &   4.25  &   6.08$\pm$ 0.43  &    0.82$\pm$   0.11  &  0.46$\pm$   0.02  &  22.78$\pm$  0.05  &   0.719 \\
ero\_141559.96+521057.4  &  0.98$\pm$0.14  &   1  &     secure  &  22.53$\pm$ 0.03  &  18.48$\pm$ 0.07  &   4.05  &  15.23$\pm$15.24  &   11.11$\pm$   4.07  &  0.83$\pm$   0.06  &  21.42$\pm$  0.40  &   0.756 \\
ero\_141600.38+520846.2  &  1.71$\pm$0.19  &  2b  &     secure  &  24.26$\pm$ 0.00  &  20.08$\pm$ 0.68  &   4.17  &   4.67$\pm$ 0.35  &    0.28$\pm$   0.14  &  0.37$\pm$   0.04  &  23.95$\pm$  0.08  &   0.749 \\
ero\_141600.97+520908.0  &  1.34$\pm$0.16  &  2b  &     secure  &  24.40$\pm$ 0.02  &  20.21$\pm$ 0.68  &   4.19  &   3.13$\pm$ 0.26  &    0.31$\pm$   0.20  &  0.30$\pm$   0.04  &  24.24$\pm$  0.07  &   0.711 \\
ero\_141601.22+521101.4$^{*}$   &  0.94$\pm$0.14  &  2c  &  un-secure  &  23.44$\pm$ 0.03  &  18.97$\pm$ 0.02  &   4.47  &   5.14$\pm$ 0.91  &    2.15$\pm$   0.47  &  0.32$\pm$   0.03  &  23.13$\pm$  0.11  &   0.758 \\
ero\_141603.15+521140.8  &  1.05$\pm$0.14  &   3  &     secure  &  24.10$\pm$ 0.00  &  19.20$\pm$ 0.01  &   4.90  &   4.40$\pm$ 7.03  &    0.04$\pm$   0.15  &  0.71$\pm$   0.10  &  24.21$\pm$  0.00  &   0.750 \\
ero\_141604.29+520925.9  &  2.74$\pm$0.26  &   1  &     secure  &  24.57$\pm$ 0.02  &  20.29$\pm$ 0.05  &   4.28  &   0.99$\pm$ 0.12  &    0.72$\pm$   0.60  &  0.46$\pm$   0.10  &  24.32$\pm$  0.06  &   0.720 \\
ero\_141605.16+520903.5  &  0.90$\pm$0.13  &   4  &     secure  &  24.52$\pm$ 0.01  &  20.41$\pm$ 0.44  &   4.11  &   6.81$\pm$ 3.76  &    3.08$\pm$   1.64  &  0.17$\pm$   0.04  &  24.02$\pm$  0.31  &   0.743 \\
ero\_141608.87+521132.3  &  2.35$\pm$0.23  &   3  &  un-secure  &  23.92$\pm$ 0.03  &  19.87$\pm$ 0.82  &   4.05  &  38.01$\pm$98.69  &   11.70$\pm$  10.40  &  0.17$\pm$   0.03  &  22.72$\pm$  1.01  &   0.725 \\
ero\_141611.77+521316.9  &  1.28$\pm$0.16  &   1  &     secure  &  23.95$\pm$ 0.72  &  19.91$\pm$ 0.04  &   4.03  &   0.34$\pm$82.83  &    0.02$\pm$  99.00  &  0.61$\pm$  99.00  &  24.67$\pm$ 99.00  &   1.052 \\
ero\_141617.81+521413.4$^{\#}$  &  1.75$\pm$0.19  &   1  &     secure  &  24.14$\pm$ 0.31  &  19.85$\pm$ 0.93  &   4.29  &   0.01$\pm$ 0.06  &    7.03$\pm$  17.23  &  0.79$\pm$  11.86  &  23.90$\pm$  0.00  &   0.781 \\
ero\_141620.23+521317.2  &  2.20$\pm$0.22  &  2b  &     secure  &  25.32$\pm$ 0.01  &  21.17$\pm$ 0.49  &   4.15  &   3.97$\pm$ 0.44  &    0.08$\pm$   0.29  &  0.21$\pm$   0.05  &  24.83$\pm$  0.12  &   0.723 \\
ero\_141628.33+521419.3  &  1.80$\pm$0.20  &  2b  &     secure  &  24.27$\pm$ 0.01  &  20.24$\pm$ 0.19  &   4.03  &   6.52$\pm$ 1.84  &    1.99$\pm$   0.72  &  0.19$\pm$   0.04  &  23.93$\pm$  0.18  &   0.704 \\
ero\_141629.52+521507.6  &  1.27$\pm$0.16  &  2a  &     secure  &  24.86$\pm$ 0.02  &  20.55$\pm$ 0.81  &   4.31  &   1.24$\pm$ 0.17  &    0.85$\pm$   0.71  &  0.19$\pm$   0.17  &  24.57$\pm$  0.07  &   0.743 \\
ero\_141631.87+521739.0  &  0.71$\pm$0.12  &   3  &     secure  &  23.98$\pm$ 0.02  &  19.92$\pm$ 0.05  &   4.05  &   4.73$\pm$ 0.41  &    0.64$\pm$   0.21  &  0.23$\pm$   0.03  &  23.79$\pm$  0.07  &   0.706 \\
ero\_141633.36+521639.7  &  2.12$\pm$0.22  &   1  &     secure  &  24.76$\pm$ 0.09  &  19.93$\pm$ 0.94  &   4.83  &   0.52$\pm$ 0.12  &    0.69$\pm$   1.42  &  1.00$\pm$   0.39  &  24.60$\pm$  0.06  &   0.711 \\
ero\_141634.23+521722.7  &  1.23$\pm$0.16  &  2c  &  un-secure  &  23.55$\pm$ 0.01  &  19.48$\pm$ 0.45  &   4.08  &   8.44$\pm$ 1.74  &    1.85$\pm$   0.39  &  0.39$\pm$   0.04  &  23.03$\pm$  0.15  &   0.712 \\
ero\_141634.75+521728.8  &  2.54$\pm$0.25  &   1  &     secure  &  24.00$\pm$ 0.03  &  19.72$\pm$ 0.96  &   4.27  &   1.14$\pm$ 0.24  &    3.22$\pm$   1.86  &  0.53$\pm$   0.11  &  24.01$\pm$  0.11  &   0.738 \\
ero\_141635.74+521451.0  &  1.40$\pm$0.17  &   1  &     secure  &  23.06$\pm$ 0.03  &  18.87$\pm$ 0.08  &   4.19  &   1.24$\pm$ 0.10  &    2.99$\pm$   0.60  &  0.54$\pm$   0.05  &  22.69$\pm$  0.03  &   0.709 \\
ero\_141636.32+521805.9  &  1.22$\pm$0.16  &   4  &  un-secure  &  24.86$\pm$ 0.01  &  20.62$\pm$ 0.54  &   4.25  &   2.76$\pm$ 2.60  &    4.08$\pm$   3.73  &  0.99$\pm$   0.34  &  24.35$\pm$  0.51  &   0.729 \\
ero\_141636.41+521449.0  &  1.65$\pm$0.19  &  2b  &     secure  &  24.15$\pm$ 0.02  &  20.01$\pm$ 0.27  &   4.14  &   4.87$\pm$ 0.50  &    0.51$\pm$   0.14  &  0.25$\pm$   0.03  &  23.81$\pm$  0.07  &   0.717 \\
ero\_141636.67+521806.8  &  1.35$\pm$0.16  &  2a  &     secure  &  23.76$\pm$ 0.03  &  19.13$\pm$ 0.67  &   4.63  &   5.32$\pm$ 1.00  &    1.88$\pm$   0.44  &  0.40$\pm$   0.04  &  23.33$\pm$  0.12  &   0.731 \\
ero\_141639.57+521810.2  &  1.71$\pm$0.19  &  2b  &     secure  &  23.97$\pm$ 0.01  &  19.88$\pm$ 0.13  &   4.09  &   5.98$\pm$ 0.24  &    0.16$\pm$   0.09  &  0.17$\pm$   0.02  &  23.65$\pm$  0.06  &   0.722 \\
ero\_141642.05+521601.7  &  1.49$\pm$0.17  &  2b  &     secure  &  24.08$\pm$ 0.00  &  19.85$\pm$ 0.26  &   4.22  &   9.08$\pm$ 0.68  &    0.26$\pm$   0.11  &  0.23$\pm$   0.02  &  23.69$\pm$  0.07  &   0.707 \\
ero\_141642.20+521641.7  &  0.92$\pm$0.13  &  2b  &     secure  &  25.17$\pm$ 0.01  &  20.25$\pm$ 0.82  &   4.93  &   2.24$\pm$ 0.32  &    0.76$\pm$   0.37  &  0.48$\pm$   0.09  &  24.55$\pm$  0.10  &   0.801 \\
ero\_141642.25+521820.2  &  0.90$\pm$0.13  &  2c  &  un-secure  &  25.28$\pm$ 0.01  &  20.93$\pm$ 0.56  &   4.35  &   1.53$\pm$ 0.35  &    1.47$\pm$   1.09  &  0.62$\pm$   0.17  &  24.72$\pm$  0.14  &   0.734 \\
ero\_141643.79+521915.4  &  1.82$\pm$0.20  &   1  &     secure  &  23.96$\pm$ 0.03  &  19.77$\pm$ 0.22  &   4.18  &   2.84$\pm$ 0.49  &    2.34$\pm$   0.72  &  0.50$\pm$   0.07  &  23.56$\pm$  0.11  &   0.719 \\
ero\_141644.29+521828.4  &  1.50$\pm$0.17  &  2a  &     secure  &  23.85$\pm$ 0.03  &  19.66$\pm$ 0.89  &   4.18  &   3.85$\pm$ 1.81  &    4.92$\pm$   2.19  &  0.44$\pm$   0.08  &  23.34$\pm$  0.23  &   0.749 \\
ero\_141645.18+521650.7  &  0.67$\pm$0.12  &  2a  &     secure  &  22.02$\pm$ 0.03  &  17.90$\pm$ 0.51  &   4.12  &  12.25$\pm$ 1.79  &    3.04$\pm$   0.32  &  0.42$\pm$   0.01  &  21.07$\pm$  0.10  &   0.721 \\
ero\_141646.01+521833.1$^{\#}$  &  1.47$\pm$0.17  &   1  &     secure  &  24.18$\pm$ 0.30  &  20.02$\pm$ 0.85  &   4.15  &   0.01$\pm$ 0.26  &   12.13$\pm$ 114.81  &  1.00$\pm$   2.45  &  23.69$\pm$  6.93  &   0.710 \\
ero\_141650.05+522119.6  &  1.29$\pm$0.16  &  2c  &  un-secure  &  22.98$\pm$ 0.03  &  18.93$\pm$ 0.59  &   4.05  &   3.44$\pm$ 0.10  &    0.33$\pm$   0.06  &  0.43$\pm$   0.02  &  22.78$\pm$  0.03  &   0.728 \\
ero\_141650.65+521825.5$^{\#}$  &  1.67$\pm$0.19  &   4  &  un-secure  &  24.45$\pm$ 0.44  &  20.07$\pm$ 0.49  &   4.38  &   0.01$\pm$ 0.03  &    8.94$\pm$   7.39  &  0.99$\pm$   3.79  &  24.13$\pm$  0.00  &   0.736 \\
ero\_141659.11+521920.7  &  1.25$\pm$0.16  &  2a  &     secure  &  23.09$\pm$ 0.03  &  18.82$\pm$ 0.03  &   4.28  &   3.22$\pm$ 0.41  &    1.64$\pm$   0.41  &  0.41$\pm$   0.05  &  23.20$\pm$  0.08  &   1.049 \\
ero\_141705.04+522306.0  &  0.37$\pm$0.10  &   1  &     secure  &  22.86$\pm$ 0.03  &  18.73$\pm$ 0.02  &   4.13  &  29.96$\pm$59.30  &   14.39$\pm$   7.56  &  0.65$\pm$   0.06  &  21.29$\pm$  0.74  &   0.708 \\
ero\_141705.15+522212.4  &  0.99$\pm$0.14  &   1  &     secure  &  22.26$\pm$ 0.03  &  18.14$\pm$ 0.03  &   4.12  &   4.48$\pm$ 0.72  &    4.80$\pm$   0.64  &  0.87$\pm$   0.04  &  21.58$\pm$  0.08  &   0.698 \\
ero\_141706.84+522225.9  &  0.90$\pm$0.13  &  2b  &     secure  &  22.36$\pm$ 0.04  &  18.18$\pm$ 0.02  &   4.18  &   5.95$\pm$ 0.42  &    2.35$\pm$   0.20  &  0.23$\pm$   0.01  &  21.97$\pm$  0.04  &   0.767 \\
ero\_141709.19+522135.3  &  1.34$\pm$0.16  &   3  &  un-secure  &  23.58$\pm$ 0.07  &  19.57$\pm$ 0.56  &   4.01  &   3.42$\pm$ 0.52  &    1.84$\pm$   0.36  &  0.96$\pm$   0.09  &  23.16$\pm$  0.10  &   0.721 \\
ero\_141709.70+522449.3  &  1.55$\pm$0.18  &  2b  &     secure  &  23.98$\pm$ 0.02  &  19.32$\pm$ 0.68  &   4.65  &   4.93$\pm$ 0.35  &    0.34$\pm$   0.15  &  0.20$\pm$   0.03  &  23.84$\pm$  0.07  &   0.729 \\
ero\_141710.62+522109.6  &  2.01$\pm$0.21  &   1  &     secure  &  23.75$\pm$ 0.23  &  19.63$\pm$ 0.55  &   4.12  &   0.17$\pm$ 0.17  &   12.29$\pm$  16.25  &  0.46$\pm$   0.22  &  23.37$\pm$  0.35  &   0.721 \\
ero\_141711.42+522111.4  &  1.21$\pm$0.15  &   1  &     secure  &  24.48$\pm$ 0.11  &  20.36$\pm$ 0.51  &   4.12  &   0.67$\pm$ 0.14  &    3.10$\pm$   2.39  &  0.87$\pm$   0.27  &  24.18$\pm$  0.10  &   0.725 \\
ero\_141714.03+522332.6  &  1.01$\pm$0.14  &   3  &     secure  &  22.81$\pm$ 0.03  &  18.65$\pm$ 0.15  &   4.16  &   2.97$\pm$ 0.12  &    1.15$\pm$   0.11  &  0.51$\pm$   0.02  &  22.53$\pm$  0.02  &   0.694 \\
ero\_141714.14+522538.9  &  1.79$\pm$0.20  &   1  &     secure  &  24.41$\pm$ 0.03  &  19.81$\pm$ 0.14  &   4.60  &   1.44$\pm$ 2.10  &   10.87$\pm$  11.18  &  0.46$\pm$   0.20  &  24.01$\pm$  0.33  &   0.747 \\
ero\_141715.09+522142.6  &  2.65$\pm$0.26  &   1  &     secure  &  24.73$\pm$ 0.28  &  20.06$\pm$ 0.58  &   4.67  &   3.45$\pm$ 4.48  &   10.88$\pm$  11.37  &  0.06$\pm$   0.07  &  24.18$\pm$  0.45  &   0.723 \\
ero\_141716.69+522549.1  &  1.91$\pm$0.20  &   4  &  un-secure  &  24.71$\pm$ 0.01  &  20.11$\pm$ 0.66  &   4.60  &   0.87$\pm$ 0.21  &    2.19$\pm$   1.97  &  0.48$\pm$   0.22  &  24.66$\pm$  0.00  &   1.046 \\
ero\_141721.02+522343.5  &  1.43$\pm$0.17  &   1  &     secure  &  23.44$\pm$ 0.03  &  19.29$\pm$ 0.79  &   4.16  &   1.18$\pm$ 0.10  &    2.06$\pm$   0.56  &  0.46$\pm$   0.06  &  23.19$\pm$  0.03  &   0.703 \\
ero\_141722.55+522345.6  &  1.41$\pm$0.17  &   1  &     secure  &  22.33$\pm$ 0.05  &  18.13$\pm$ 0.12  &   4.20  &   0.78$\pm$ 0.06  &    5.50$\pm$   0.93  &  0.86$\pm$   0.05  &  22.01$\pm$  0.02  &   0.704 \\
ero\_141723.61+522555.2$^{*}$   &  1.28$\pm$0.16  &   3  &     secure  &  23.45$\pm$ 0.02  &  18.94$\pm$ 0.01  &   4.51  &  11.01$\pm$ 0.63  &    0.23$\pm$   0.06  &  0.32$\pm$   0.02  &  23.11$\pm$  0.06  &   0.748 \\
ero\_141726.68+522415.1  &  1.12$\pm$0.15  &  2a  &     secure  &  24.12$\pm$ 0.29  &  19.91$\pm$ 0.09  &   4.21  &   3.77$\pm$ 1.21  &    3.22$\pm$   1.33  &  0.29$\pm$   0.06  &  23.75$\pm$  0.18  &   0.711 \\
ero\_141726.95+522449.6$^{*}$   &  1.17$\pm$0.15  &  2b  &     secure  &  22.99$\pm$ 0.00  &  18.76$\pm$ 0.02  &   4.23  &  11.26$\pm$ 0.60  &    0.13$\pm$   0.08  &  0.12$\pm$   0.01  &  23.34$\pm$  0.00  &   0.765 \\
ero\_141727.73+522411.6  &  1.47$\pm$0.17  &  2a  &     secure  &  23.92$\pm$ 0.03  &  19.66$\pm$ 0.30  &   4.26  &   1.74$\pm$ 0.15  &    1.33$\pm$   0.44  &  0.21$\pm$   0.06  &  23.66$\pm$  0.04  &   0.717 \\
ero\_141728.35+522606.0  &  2.35$\pm$0.23  &   1  &     secure  &  24.80$\pm$ 0.03  &  19.75$\pm$ 0.17  &   5.04  &   1.35$\pm$ 0.30  &    1.54$\pm$   1.28  &  0.18$\pm$   0.19  &  24.79$\pm$  0.08  &   0.714 \\
ero\_141730.64+522823.8  &  1.27$\pm$0.16  &   3  &  un-secure  &  24.84$\pm$ 0.02  &  20.77$\pm$ 0.30  &   4.07  &   0.88$\pm$ 0.32  &    3.52$\pm$   3.63  &  0.79$\pm$   0.26  &  24.59$\pm$  0.20  &   0.720 \\
ero\_141731.31+522507.0  &  2.90$\pm$0.27  &   3  &  un-secure  &  24.81$\pm$ 0.50  &  20.79$\pm$ 0.52  &   4.02  &  10.21$\pm$20.32  &    5.34$\pm$   5.76  &  0.37$\pm$   0.14  &  23.91$\pm$  1.04  &   0.740 \\
ero\_141735.49+522554.4  &  1.45$\pm$0.17  &   1  &     secure  &  23.77$\pm$ 0.13  &  18.92$\pm$ 0.02  &   4.84  &   1.42$\pm$ 0.21  &    3.16$\pm$   1.22  &  0.62$\pm$   0.07  &  23.48$\pm$  0.08  &   0.778 \\
ero\_141739.07+522843.8  &  0.91$\pm$0.13  &  2a  &     secure  &  23.75$\pm$ 0.03  &  19.50$\pm$ 0.03  &   4.25  &   2.05$\pm$ 0.17  &    1.90$\pm$   0.46  &  0.60$\pm$   0.06  &  23.25$\pm$  0.06  &   0.720 \\
ero\_141740.22+522905.9  &  1.65$\pm$0.19  &   1  &     secure  &  24.74$\pm$ 0.03  &  19.49$\pm$ 0.03  &   5.25  &   5.16$\pm$ 2.19  &    4.28$\pm$   1.36  &  0.65$\pm$   0.08  &  23.88$\pm$  0.23  &   0.930 \\
ero\_141740.84+522649.4  &  1.75$\pm$0.19  &   1  &     secure  &  24.44$\pm$ 0.02  &  19.74$\pm$ 0.03  &   4.71  &   2.24$\pm$ 7.76  &   19.99$\pm$  46.59  &  1.00$\pm$   0.35  &  23.57$\pm$  1.25  &   0.702 \\
ero\_141742.38+523034.4  &  1.08$\pm$0.15  &  2a  &     secure  &  24.25$\pm$ 0.02  &  19.66$\pm$ 0.90  &   4.60  &  24.38$\pm$59.47  &    8.71$\pm$   8.13  &  0.26$\pm$   0.07  &  23.09$\pm$  1.08  &   0.725 \\
ero\_141742.39+522811.5  &  0.99$\pm$0.14  &  2b  &     secure  &  23.61$\pm$ 0.00  &  19.36$\pm$ 0.03  &   4.26  &   9.30$\pm$ 0.90  &    0.66$\pm$   0.13  &  0.29$\pm$   0.02  &  23.13$\pm$  0.07  &   0.713 \\
ero\_141744.08+522631.2$^{\#}$  &  1.99$\pm$0.21  &   1  &     secure  &  25.07$\pm$ 0.36  &  20.52$\pm$ 0.06  &   4.55  &   0.05$\pm$ 6.60  &    3.15$\pm$ 333.19  &  0.78$\pm$   9.04  &  24.86$\pm$ 10.73  &   0.721 \\
ero\_141744.28+522925.0  &  0.75$\pm$0.12  &   1  &     secure  &  25.33$\pm$ 0.03  &  20.71$\pm$ 0.01  &   4.62  &   0.91$\pm$ 0.23  &    3.57$\pm$   2.43  &  0.67$\pm$   0.18  &  25.15$\pm$  0.14  &   0.936 \\
ero\_141745.13+523045.9  &  1.40$\pm$0.17  &   4  &  un-secure  &  25.02$\pm$ 0.00  &  20.83$\pm$ 0.52  &   4.19  &   3.50$\pm$ 0.72  &    0.99$\pm$   0.34  &  0.50$\pm$   0.09  &  24.35$\pm$  0.14  &   0.733 \\
ero\_141746.71+522857.8  &  1.59$\pm$0.18  &  2b  &     secure  &  25.43$\pm$ 0.02  &  21.18$\pm$ 0.02  &   4.26  &   2.58$\pm$ 0.22  &    0.29$\pm$   0.19  &  0.53$\pm$   0.05  &  25.19$\pm$  0.07  &   0.924 \\
ero\_141749.11+522759.6  &  0.93$\pm$0.14  &   1  &     secure  &  24.26$\pm$ 0.06  &  20.22$\pm$ 0.01  &   4.04  &   0.91$\pm$ 0.09  &    2.15$\pm$   0.68  &  0.52$\pm$   0.06  &  24.15$\pm$  0.03  &   1.071 \\
ero\_141749.24+522811.0  &  1.08$\pm$0.15  &   3  &     secure  &  23.11$\pm$ 0.16  &  19.04$\pm$ 0.03  &   4.08  &   1.65$\pm$ 0.10  &    5.70$\pm$   0.69  &  0.39$\pm$   0.04  &  23.11$\pm$  0.00  &   1.255 \\
ero\_141749.59+522806.2  &  1.51$\pm$0.18  &  2b  &     secure  &  24.34$\pm$ 0.03  &  20.16$\pm$ 0.03  &   4.19  &   4.46$\pm$ 0.30  &    0.62$\pm$   0.16  &  0.28$\pm$   0.02  &  24.11$\pm$  0.05  &   1.371 \\
ero\_141751.29+523040.3  &  1.51$\pm$0.18  &  2a  &     secure  &  23.84$\pm$ 0.03  &  19.65$\pm$ 0.81  &   4.19  &   2.73$\pm$ 0.26  &    1.15$\pm$   0.21  &  0.73$\pm$   0.06  &  23.49$\pm$  0.06  &   0.710 \\
ero\_141751.38+523049.8  &  1.54$\pm$0.18  &   3  &     secure  &  24.88$\pm$ 0.01  &  19.79$\pm$ 0.47  &   5.09  &   4.94$\pm$ 0.34  &    0.16$\pm$   0.15  &  0.23$\pm$   0.04  &  24.34$\pm$  0.09  &   0.712 \\
ero\_141751.76+523136.8  &  1.33$\pm$0.16  &  2a  &     secure  &  23.26$\pm$ 0.03  &  18.54$\pm$ 0.08  &   4.72  &   3.16$\pm$ 0.44  &    2.48$\pm$   0.50  &  0.73$\pm$   0.06  &  22.93$\pm$  0.09  &   0.716 \\
ero\_141754.50+523023.4  &  4.60$\pm$0.39  &   1  &     secure  &  24.93$\pm$ 0.23  &  20.46$\pm$ 0.46  &   4.47  &   0.35$\pm$ 0.19  &    6.95$\pm$  12.99  &  0.78$\pm$   0.51  &  24.61$\pm$  0.37  &   0.723 \\
ero\_141755.44+522928.5  &  2.98$\pm$0.28  &   4  &  un-secure  &  24.70$\pm$ 0.04  &  20.43$\pm$ 0.62  &   4.27  &   0.81$\pm$ 0.11  &    1.26$\pm$   1.10  &  0.22$\pm$   0.16  &  24.47$\pm$  0.07  &   0.708 \\
ero\_141756.74+523157.2  &  1.32$\pm$0.16  &  2c  &  un-secure  &  23.04$\pm$ 0.03  &  18.64$\pm$ 0.14  &   4.40  &   4.92$\pm$ 0.51  &    1.33$\pm$   0.17  &  0.65$\pm$   0.04  &  22.87$\pm$  0.07  &   0.726 \\
ero\_141756.91+523118.1  &  1.48$\pm$0.17  &   4  &  un-secure  &  24.29$\pm$ 0.01  &  20.24$\pm$ 0.47  &   4.04  &   2.72$\pm$ 3.11  &    6.12$\pm$   6.92  &  0.54$\pm$   0.20  &  24.31$\pm$  0.54  &   0.718 \\
ero\_141757.15+523242.6  &  1.61$\pm$0.18  &   1  &     secure  &  24.27$\pm$ 0.07  &  19.69$\pm$ 0.16  &   4.58  &   0.80$\pm$ 0.29  &    6.34$\pm$   5.66  &  0.47$\pm$   0.20  &  24.14$\pm$  0.14  &   0.721 \\
ero\_141757.27+523224.5$^{\#}$  &  1.34$\pm$0.16  &   1  &     secure  &  24.29$\pm$ 0.60  &  19.49$\pm$ 0.71  &   4.80  &   0.01$\pm$17.87  &    4.81$\pm$3912.27  &  0.83$\pm$ 119.26  &  24.00$\pm$738.06  &   0.719 \\
ero\_141757.66+522910.3  &  1.33$\pm$0.16  &  2b  &     secure  &  24.04$\pm$ 0.00  &  19.85$\pm$ 0.00  &   4.19  &   4.48$\pm$ 1.55  &    2.30$\pm$   0.86  &  0.56$\pm$   0.10  &  23.68$\pm$  0.22  &   0.739 \\
ero\_141800.87+523203.0  &  1.59$\pm$0.18  &  2a  &     secure  &  23.93$\pm$ 0.03  &  19.07$\pm$ 0.03  &   4.86  &   2.07$\pm$ 0.36  &    2.96$\pm$   1.00  &  0.81$\pm$   0.08  &  23.42$\pm$  0.11  &   0.720 \\
ero\_141802.03+523015.5  &  1.64$\pm$0.18  &  2b  &     secure  &  24.05$\pm$ 0.03  &  19.70$\pm$ 0.08  &   4.35  &   2.69$\pm$ 0.17  &    0.07$\pm$   0.32  &  0.29$\pm$   0.05  &  24.41$\pm$  0.08  &   0.711 \\
ero\_141802.57+523251.8  &  1.73$\pm$0.19  &   4  &  un-secure  &  25.00$\pm$ 0.01  &  19.86$\pm$ 0.59  &   5.14  &   5.82$\pm$ 2.04  &    1.56$\pm$   0.78  &  0.18$\pm$   0.05  &  24.52$\pm$  0.22  &   0.719 \\
ero\_141803.00+523033.9  &  1.35$\pm$0.16  &   4  &  un-secure  &  24.97$\pm$ 0.01  &  20.92$\pm$ 0.49  &   4.05  &   1.21$\pm$ 0.34  &    0.11$\pm$   1.06  &  0.85$\pm$   0.21  &  24.81$\pm$  0.08  &   0.736 \\
ero\_141803.34+523228.4  &  1.10$\pm$0.15  &   1  &     secure  &  24.62$\pm$ 0.08  &  19.97$\pm$ 0.52  &   4.65  &   1.21$\pm$ 0.54  &    7.08$\pm$   6.38  &  0.08$\pm$   0.12  &  24.29$\pm$  0.18  &   0.717 \\
ero\_141809.26+523112.5  &  0.84$\pm$0.13  &  2b  &     secure  &  23.46$\pm$ 0.02  &  19.19$\pm$ 0.01  &   4.27  &   6.68$\pm$ 0.54  &    1.10$\pm$   0.17  &  0.20$\pm$   0.01  &  22.94$\pm$  0.05  &   0.726 \\

%

\enddata
\end{deluxetable}

\clearpage

\onecolumn
\begin{deluxetable}{lcrrcccccccc}
\tablewidth{0pt}
\tablecaption{Results of mass and age analysis : (2): formation redshift z$_\mathrm{f}$, (3): exponential decay time $\tau$, (4): optical depth A$_\mathrm{v}$, (5): metallicity Z, (6): luminosity distance D$_\mathrm{L}$, (7): total stellar mass M$_\mathrm{tot}$, (8): age, (9) $\chi^{2}$ }
\label{tab2}

\tablehead{

\colhead{Name} & \colhead{z$_\mathrm{f}$} & \colhead{$\tau$} &\colhead{A$_\mathrm{v}$} & \colhead{Z} &\colhead{D$_\mathrm{L}$} & \colhead{M$_\mathrm{tot}$}      &\colhead{age}        &  \colhead{$\chi^{2}$}\\
\colhead{}     & \colhead{}              & \colhead{Gyr}    &\colhead{}              &\colhead{}   &\colhead{/10$^{3}$Mpc}         & \colhead{/10$^{10}$M$_{\odot}$} &\colhead{/10$^{8}$yr} &  \colhead{}         \\
\colhead{(1)}  & \colhead{(2)}           & \colhead{(3)}    &\colhead{(4)}           &\colhead{(5)}&\colhead{(6)}                 & \colhead{(7)}                  &\colhead{(8)}         &  \colhead{(9)} }
\startdata
ero\_141520.43+520300.8  &  3  &   7.00  &   2.0  &   0.004  &      5.92$^{+1.05}_{-1.09}$  &   2.216  &  18.000  &     0.45 \\
ero\_141521.67+520358.0  &  3  &   0.50  &   1.0  &   0.008  &      6.40$^{+1.10}_{-1.14}$  &   3.690  &  21.000  &     1.04 \\
ero\_141521.72+520354.2  &  4  &   0.50  &   1.0  &   0.020  &      6.94$^{+1.16}_{-1.20}$  &  11.596  &  23.000  &     3.13 \\
ero\_141524.80+520419.0  &  5  &   1.00  &   0.6  &   0.050  &      6.20$^{+1.08}_{-1.12}$  &   8.735  &  45.000  &     5.92 \\
ero\_141526.12+520555.9  &  3  &   0.05  &   0.6  &   0.050  &     11.95$^{+1.65}_{-1.69}$  &  32.103  &  10.150  &    28.18 \\
ero\_141526.29+520417.5  &  8  &   4.00  &   0.6  &   0.004  &     18.82$^{+2.27}_{-2.31}$  &   2.575  &  21.000  &    39.75 \\
ero\_141526.54+520258.1  &  3  &   0.05  &   0.6  &   0.050  &     13.19$^{+1.76}_{-1.80}$  &   8.471  &   5.088  &    16.25 \\
ero\_141526.65+520405.0  &  3  &   0.10  &   2.0  &   0.020  &      5.84$^{+1.04}_{-1.08}$  &   0.781  &   4.535  &    12.43 \\
ero\_141528.89+520415.9  &  3  &   0.10  &   0.6  &   0.050  &     15.49$^{+1.97}_{-2.01}$  &   6.118  &   4.535  &     1.33 \\
ero\_141530.47+520504.3  &  3  &   0.50  &   2.0  &   0.020  &      4.11$^{+0.85}_{-0.89}$  &  12.818  &  12.780  &     1.78 \\
ero\_141531.91+520339.0  &  3  &   0.05  &   1.0  &   0.050  &     11.13$^{+1.57}_{-1.61}$  &   6.340  &   3.210  &     1.53 \\
ero\_141532.19+520438.6  &  3  &   0.10  &   0.6  &   0.050  &     10.02$^{+1.46}_{-1.50}$  &   5.981  &   6.405  &     3.36 \\
ero\_141534.63+520303.0  &  6  &   0.50  &   1.0  &   0.008  &     22.09$^{+2.55}_{-2.59}$  &  16.444  &  12.780  &     7.02 \\
ero\_141541.34+520749.7  &  4  &   0.05  &   1.0  &   0.050  &      7.39$^{+1.20}_{-1.24}$  &   2.473  &   1.805  &    42.22 \\
ero\_141541.93+520639.0  &  3  &   0.05  &   0.6  &   0.050  &     12.68$^{+1.72}_{-1.76}$  &  23.354  &   5.088  &    20.08 \\
ero\_141542.14+520643.8  &  3  &   0.10  &   2.0  &   0.050  &      6.08$^{+1.07}_{-1.11}$  &   2.988  &   4.042  &     3.00 \\
ero\_141544.23+520731.9  &  8  &   1.00  &   0.6  &   0.008  &      7.35$^{+1.20}_{-1.24}$  &   7.365  &  47.500  &     1.79 \\
ero\_141546.60+520921.1  &  4  &   0.50  &   0.6  &   0.004  &     13.42$^{+1.78}_{-1.82}$  &   4.458  &  16.090  &     8.64 \\
ero\_141547.57+520653.5  &  6  &   1.00  &   0.6  &   0.004  &     11.13$^{+1.57}_{-1.61}$  &   9.129  &  30.000  &     8.72 \\
ero\_141547.81+520912.0  &  4  &   0.05  &   0.6  &   0.050  &     13.94$^{+1.83}_{-1.87}$  &  12.119  &   2.861  &     5.97 \\
ero\_141550.16+520706.3  &  3  &   0.10  &   0.6  &   0.050  &     11.81$^{+1.64}_{-1.67}$  &  18.843  &   7.187  &     8.84 \\
ero\_141551.31+520954.3  &  3  &   0.10  &   2.0  &   0.020  &      6.36$^{+1.10}_{-1.14}$  &   2.688  &   4.535  &     5.68 \\
ero\_141551.59+521030.1  &  3  &   0.50  &   1.0  &   0.020  &      7.14$^{+1.18}_{-1.22}$  &   5.610  &  16.800  &     3.00 \\
ero\_141552.93+520701.4  &  3  &   0.10  &   0.6  &   0.050  &     11.31$^{+1.59}_{-1.63}$  &  10.474  &   5.709  &     7.64 \\
ero\_141552.95+520739.5  &  5  &   1.00  &   1.0  &   0.050  &     13.61$^{+1.80}_{-1.84}$  &  18.458  &  21.000  &     1.20 \\
ero\_141553.33+520718.4  &  4  &   0.50  &   0.6  &   0.004  &      9.36$^{+1.40}_{-1.44}$  &  10.254  &  27.500  &    12.49 \\
ero\_141557.80+521052.2  &  3  &   0.05  &   2.0  &   0.004  &      5.76$^{+1.03}_{-1.07}$  &   3.210  &   3.602  &     2.41 \\
ero\_141559.96+521057.4  &  3  &   0.05  &   1.0  &   0.050  &      6.49$^{+1.11}_{-1.15}$  &   8.592  &   6.405  &     9.32 \\
ero\_141600.38+520846.2  &  5  &   1.00  &   0.6  &   0.004  &     12.82$^{+1.73}_{-1.77}$  &   7.748  &  23.000  &     0.47 \\
ero\_141600.97+520908.0  &  3  &   0.10  &   0.6  &   0.050  &      9.53$^{+1.42}_{-1.46}$  &   3.119  &   5.088  &     0.65 \\
ero\_141601.22+521101.4  &  3  &   0.50  &   1.0  &   0.020  &      6.16$^{+1.08}_{-1.12}$  &   8.560  &  24.000  &     0.67 \\
ero\_141603.15+521140.8  &  4  &   0.50  &   0.6  &   0.050  &      7.06$^{+1.17}_{-1.21}$  &   8.658  &  27.500  &     1.62 \\
ero\_141604.29+520925.9  &  7  &   1.00  &   0.6  &   0.004  &     22.79$^{+2.61}_{-2.65}$  &   8.224  &  14.340  &     8.32 \\
ero\_141605.16+520903.5  & ... &   ...   &   ...  &   ...    &      5.76$^{+1.03}_{-1.07}$  &    ...   &   ...    &     ...  \\
ero\_141608.87+521132.3  &  4  &   0.05  &   0.6  &   0.008  &     18.87$^{+2.27}_{-2.31}$  &   7.863  &   2.550  &     2.36 \\
ero\_141611.77+521316.9  &  3  &   0.10  &   0.6  &   0.050  &      9.01$^{+1.37}_{-1.41}$  &   5.529  &   7.187  &     4.97 \\
ero\_141617.81+521413.4  &  3  &   0.05  &   0.6  &   0.050  &     13.24$^{+1.77}_{-1.81}$  &  14.111  &   4.535  &    21.89 \\
ero\_141620.23+521317.2  &  5  &   0.50  &   0.6  &   0.004  &     17.46$^{+2.15}_{-2.19}$  &   5.069  &  14.340  &     6.35 \\
ero\_141628.33+521419.3  &  3  &   0.05  &   0.6  &   0.008  &     13.70$^{+1.81}_{-1.85}$  &   4.239  &   4.042  &     3.27 \\
ero\_141629.52+521507.6  &  3  &   0.10  &   0.6  &   0.050  &      8.88$^{+1.35}_{-1.39}$  &   2.272  &   7.187  &     9.82 \\
ero\_141631.87+521739.0  &  4  &   1.00  &   0.6  &   0.050  &      4.33$^{+0.87}_{-0.91}$  &   1.683  &  40.000  &     1.17 \\
ero\_141633.36+521639.7  &  3  &   0.05  &   3.0  &   0.008  &     16.64$^{+2.07}_{-2.11}$  &  26.603  &   0.905  &     5.54 \\
ero\_141634.23+521722.7  &  3  &   0.50  &   0.6  &   0.020  &      8.54$^{+1.32}_{-1.36}$  &   5.580  &  16.090  &     3.45 \\
ero\_141634.75+521728.8  &  4  &   0.10  &   1.0  &   0.008  &     20.84$^{+2.44}_{-2.48}$  &  28.976  &   4.535  &     0.21 \\
ero\_141635.74+521451.0  &  3  &   0.05  &   0.6  &   0.050  &      9.97$^{+1.46}_{-1.50}$  &  18.914  &   6.405  &    23.77 \\
ero\_141636.32+521805.9  &  5  &   1.00  &   0.6  &   0.020  &      8.45$^{+1.31}_{-1.35}$  &   3.656  &  35.000  &     2.26 \\
ero\_141636.41+521449.0  &  3  &   0.05  &   0.6  &   0.050  &     12.27$^{+1.68}_{-1.72}$  &  11.295  &   5.088  &    14.06 \\
ero\_141636.67+521806.8  &  4  &   0.50  &   0.6  &   0.004  &      9.62$^{+1.43}_{-1.47}$  &  15.017  &  30.000  &     3.09 \\
ero\_141639.57+521810.2  &  3  &   0.10  &   0.6  &   0.004  &     12.78$^{+1.73}_{-1.76}$  &   3.340  &   4.042  &     0.28 \\
ero\_141642.05+521601.7  &  3  &   0.10  &   0.6  &   0.050  &     10.82$^{+1.54}_{-1.58}$  &   9.549  &   7.187  &     7.50 \\
ero\_141642.20+521641.7  &  3  &   0.10  &   2.0  &   0.050  &      5.92$^{+1.05}_{-1.09}$  &   1.086  &   4.535  &    12.90 \\
ero\_141642.25+521820.2  & ... &   ...   &   ...  &    ...   &      5.80$^{+1.04}_{-1.08}$  &    ...   &   ...    &   ...  \\
ero\_141643.79+521915.4  &  3  &   0.10  &   2.0  &   0.050  &     13.84$^{+1.82}_{-1.86}$  &  17.196  &   2.273  &     0.52 \\
ero\_141644.29+521828.4  &  3  &   0.05  &   0.6  &   0.008  &     10.91$^{+1.55}_{-1.59}$  &   5.097  &   5.088  &     0.30 \\
ero\_141645.18+521650.7  &  3  &   0.50  &   2.0  &   0.004  &      4.04$^{+0.84}_{-0.88}$  &  12.094  &  22.000  &     0.32 \\
ero\_141646.01+521833.1  &  3  &   0.05  &   0.6  &   0.050  &     10.64$^{+1.52}_{-1.56}$  &  14.935  &  10.150  &   125.76 \\
ero\_141650.05+522119.6  &  3  &   0.05  &   0.6  &   0.050  &      9.10$^{+1.38}_{-1.41}$  &  10.042  &   5.088  &     7.84 \\
ero\_141650.65+521825.5  &  4  &   0.10  &   0.6  &   0.004  &     12.45$^{+1.70}_{-1.73}$  &   7.379  &   9.048  &     9.01 \\
ero\_141659.11+521920.7  &  3  &   0.05  &   0.6  &   0.004  &      8.71$^{+1.34}_{-1.38}$  &   9.050  &  11.390  &     3.01 \\
ero\_141705.04+522306.0  &  3  &   2.00  &   2.0  &   0.020  &      1.98$^{+0.58}_{-0.62}$  &   2.888  &  70.000  &     3.37 \\
ero\_141705.15+522212.4  &  3  &   0.10  &   2.0  &   0.008  &      6.57$^{+1.12}_{-1.16}$  &  16.754  &   5.088  &    16.63 \\
ero\_141706.84+522225.9  &  3  &   0.10  &   2.0  &   0.020  &      5.76$^{+1.03}_{-1.07}$  &   8.713  &   4.535  &     6.68 \\
ero\_141709.19+522135.3  &  4  &   0.10  &   0.6  &   0.020  &      9.53$^{+1.42}_{-1.46}$  &   6.695  &   7.187  &     1.41 \\
ero\_141709.70+522449.3  &  4  &   0.50  &   0.6  &   0.020  &     11.41$^{+1.60}_{-1.64}$  &  17.260  &  21.000  &     1.14 \\
ero\_141710.62+522109.6  &  3  &   7.00  &   2.0  &   0.050  &     15.63$^{+1.99}_{-2.02}$  &   7.967  &   1.805  &    11.11 \\
ero\_141711.42+522111.4  &  6  &   1.00  &   0.6  &   0.008  &      8.37$^{+1.30}_{-1.34}$  &   5.265  &  40.000  &     1.65 \\
ero\_141714.03+522332.6  &  6  &   1.00  &   0.6  &   0.008  &      6.73$^{+1.14}_{-1.18}$  &  18.328  &  47.500  &     4.06 \\
ero\_141714.14+522538.9  &  3  &   0.10  &   0.6  &   0.050  &     13.61$^{+1.80}_{-1.84}$  &  16.819  &   7.187  &    14.47 \\
ero\_141715.09+522142.6  &  7  &   0.50  &   1.0  &   0.004  &     21.84$^{+2.53}_{-2.57}$  &  31.400  &  16.090  &     5.68 \\
ero\_141716.69+522549.1  &  6  &   1.00  &   1.0  &   0.004  &     14.69$^{+1.90}_{-1.94}$  &  14.913  &  23.000  &     2.68 \\
ero\_141721.02+522343.5  &  3  &   0.05  &   1.0  &   0.050  &     10.28$^{+1.49}_{-1.53}$  &   8.020  &   3.210  &     1.14 \\
ero\_141722.55+522345.6  &  3  &   0.05  &   0.6  &   0.020  &     10.06$^{+1.47}_{-1.51}$  &  28.027  &   6.405  &     3.96 \\
ero\_141723.61+522555.2  &  4  &   0.10  &   0.6  &   0.050  &      9.01$^{+1.37}_{-1.41}$  &   8.897  &   6.405  &     4.12 \\
ero\_141726.68+522415.1  &  5  &   1.00  &   1.0  &   0.004  &      7.60$^{+1.23}_{-1.27}$  &   7.882  &  40.000  &     2.70 \\
ero\_141726.95+522449.6  &  3  &   0.05  &   1.0  &   0.020  &      8.02$^{+1.27}_{-1.31}$  &  12.238  &   6.405  &     1.32 \\
ero\_141727.73+522411.6  &  3  &   0.05  &   0.6  &   0.050  &     10.60$^{+1.52}_{-1.56}$  &  16.332  &   9.048  &    77.89 \\
ero\_141728.35+522606.0  &  4  &   0.10  &   0.6  &   0.050  &     18.92$^{+2.28}_{-2.32}$  &  32.707  &   6.405  &     2.68 \\
ero\_141730.64+522823.8  &  4  &   0.05  &   1.0  &   0.050  &      8.88$^{+1.35}_{-1.39}$  &   1.711  &   3.602  &     7.15 \\
ero\_141731.31+522507.0  &  6  &   0.50  &   0.6  &   0.004  &     24.41$^{+2.75}_{-2.79}$  &   8.577  &  11.390  &    21.13 \\
ero\_141735.49+522554.4  &  6  &   0.50  &   0.6  &   0.020  &     10.46$^{+1.51}_{-1.55}$  &  47.326  &  32.500  &    13.43 \\
ero\_141739.07+522843.8  &  4  &   1.00  &   0.6  &   0.050  &      5.88$^{+1.05}_{-1.09}$  &   7.750  &  45.000  &     0.08 \\
ero\_141740.22+522905.9  &  4  &   0.05  &   0.6  &   0.020  &     12.27$^{+1.68}_{-1.72}$  &  26.014  &  20.000  &    21.29 \\
ero\_141740.84+522649.4  &  3  &   0.05  &   1.0  &   0.050  &     13.24$^{+1.77}_{-1.81}$  &  11.417  &   3.602  &     3.72 \\
ero\_141742.38+523034.4  &  3  &   0.50  &   0.6  &   0.050  &      7.27$^{+1.19}_{-1.23}$  &   5.637  &  25.000  &     5.11 \\
ero\_141742.39+522811.5  &  3  &   1.00  &   0.6  &   0.020  &      6.53$^{+1.11}_{-1.15}$  &   5.522  &  35.000  &     1.00 \\
ero\_141744.08+522631.2  &  3  &   0.10  &   0.6  &   0.050  &     15.40$^{+1.96}_{-2.00}$  &   6.521  &   5.088  &     1.80 \\
ero\_141744.28+522925.0  &  3  &   7.00  &   0.6  &   0.050  &      4.64$^{+0.91}_{-0.95}$  &   0.131  &  37.500  &    10.83 \\
ero\_141745.13+523045.9  &  4  &   0.50  &   0.6  &   0.050  &     10.02$^{+1.46}_{-1.50}$  &   4.447  &  18.000  &     1.46 \\
ero\_141746.71+522857.8  &  4  &   0.50  &   0.6  &   0.050  &     11.77$^{+1.63}_{-1.67}$  &   8.992  &  23.000  &     3.39 \\
ero\_141749.11+522759.6  &  6  &   1.00  &   1.0  &   0.050  &      6.04$^{+1.06}_{-1.10}$  &   7.208  &  50.000  &     8.95 \\
ero\_141749.24+522811.0  &  5  &   1.00  &   0.6  &   0.008  &      7.31$^{+1.20}_{-1.24}$  &  13.825  &  42.500  &     0.73 \\
ero\_141749.59+522806.2  &  5  &   0.50  &   0.6  &   0.050  &     11.00$^{+1.56}_{-1.60}$  &  48.593  &  30.000  &     1.70 \\
ero\_141751.29+523040.3  &  3  &   0.05  &   0.6  &   0.004  &     11.00$^{+1.56}_{-1.60}$  &   5.025  &   6.405  &     8.02 \\
ero\_141751.38+523049.8  &  4  &   0.10  &   0.6  &   0.050  &     11.31$^{+1.59}_{-1.63}$  &  20.260  &  16.090  &    32.35 \\
ero\_141751.76+523136.8  &  3  &   0.10  &   0.6  &   0.050  &      9.40$^{+1.41}_{-1.44}$  &  27.221  &  11.390  &    18.94 \\
ero\_141754.50+523023.4  & ... &   ...   &   ...  &   ...    &     42.29$^{+4.22}_{-4.27}$  &   ...    &   ...    &   ...  \\ 
ero\_141755.44+522928.5  &  4  &   7.00  &   0.6  &   0.004  &     25.17$^{+2.81}_{-2.85}$  &   1.144  &   3.602  &    18.40 \\
ero\_141756.74+523157.2  &  3  &   0.05  &   0.6  &   0.004  &      9.36$^{+1.40}_{-1.44}$  &  11.389  &  11.390  &     2.64 \\
ero\_141756.91+523118.1  & ... &   ...   &   ...  &   ...    &     10.73$^{+1.53}_{-1.57}$  &   ...     &   ...    &   ...  \\
ero\_141757.15+523242.6  &  3  &   0.05  &   1.0  &   0.050  &     11.90$^{+1.64}_{-1.68}$  &   8.622  &   3.602  &     2.93 \\
ero\_141757.27+523224.5  &  7  &   0.50  &   0.6  &   0.020  &      9.49$^{+1.41}_{-1.45}$  &  29.389  &  37.500  &    15.33 \\
ero\_141757.66+522910.3  &  3  &   0.05  &   0.6  &   0.004  &      9.40$^{+1.41}_{-1.44}$  &   1.768  &   3.602  &     2.56 \\
ero\_141800.87+523203.0  &  3  &   0.05  &   0.6  &   0.050  &     11.77$^{+1.63}_{-1.67}$  &  27.766  &   8.064  &    16.26 \\
ero\_141802.03+523015.5  & ... &   ...   &   ...  &   ...    &     12.18$^{+1.67}_{-1.71}$  &  ...     &   ...    &   ...  \\
ero\_141802.57+523251.8  &  4  &   0.50  &   0.6  &   0.050  &     12.96$^{+1.74}_{-1.78}$  &  12.385  &  20.000  &     2.95 \\
ero\_141803.00+523033.9  &  4  &   0.50  &   0.6  &   0.020  &      9.62$^{+1.43}_{-1.47}$  &   2.889  &  16.800  &     1.33 \\
ero\_141803.34+523228.4  &  4  &   1.00  &   1.0  &   0.020  &      7.48$^{+1.21}_{-1.25}$  &   6.106  &  32.500  &     0.39 \\
ero\_141809.26+523112.5  &  6  &   1.00  &   0.6  &   0.050  &      5.37$^{+0.99}_{-1.03}$  &  10.680  &  55.000  &     1.14 \\

\enddata
\end{deluxetable}

\end{document}